\documentclass[journal,onecolumn]{IEEEtran}
\usepackage{booktabs}
\usepackage{subfigure}
\usepackage{amsmath}                
\usepackage{amsthm}                 
\usepackage{amssymb}
\usepackage{authblk}
\usepackage{thmtools}
\usepackage{kpfonts}
\usepackage[geometry]{ifsym}
\usepackage{dsfont}
\usepackage{multirow}
\usepackage[mathscr]{euscript}
\usepackage{graphicx}
\usepackage{color}
\usepackage[inline]{enumitem}
\usepackage{framed}
\usepackage{float}
\usepackage{longtable}
\usepackage{cite}
\usepackage{enumitem}
\usepackage{caption} 
\usepackage{subcaption} 
\usepackage{circuitikz}
\usepackage{tikz}

\usetikzlibrary{patterns,decorations.pathmorphing,positioning, decorations.markings}
\usepackage{xcolor}

\usepackage[colorlinks=true, citecolor=blue, linkcolor=blue]{hyperref}       
\usepackage[font=itshape]{quoting}
\usepackage{lscape}
\usepackage{makecell}
\usepackage[section]{placeins}

\usepackage{listings}
\usepackage{comment}
\usepackage{framed} 
\usepackage{nomencl} 
\makenomenclature
\allowdisplaybreaks
\declaretheoremstyle[%
  headfont=\bfseries,%
  headpunct={:},%
  notefont=\normalfont\bfseries,%
  notebraces={--~}{},
    qed=$\blacksquare$,
]{definitionstyle}
\theoremstyle{definition}
\declaretheorem[style=definitionstyle,name=Definition]{defn}

\theoremstyle{definition}

\theoremstyle{plain}
\theoremstyle{remark}

\newcommand{\liinesbigfig}[4]{\begin{figure*}[ht!]\begin{center}\includegraphics[width=#4in]{#1}\vspace{-0.1in}\caption{#2}\label{#3}\end{center}\vspace{-0.2in}\end{figure*}}

\usepackage{nicematrix}
\NiceMatrixOptions{
code-for-first-row = \color{blue} ,
code-for-last-row = \color{blue} ,
code-for-first-col = \color{blue} ,
code-for-last-col = \color{blue}
}

\begin{document}
%
\title{Convergent Anthropocene Systems-of-Systems:
Overcoming the Limitations of System Dynamics
with Hetero-functional Graph Theory}

\author[1]{Mohammad Mahdi Naderi\textsuperscript{*}}
\author[1]{Megan Harris}
\author[2]{Ehsanoddin Ghorbanichemazkati}
\author[1]{John C. Little}
\author[2]{Amro M. Farid}

\affil[1]{Department of Civil and Environmental Engineering, Virginia Tech, Blacksburg, Virginia 24061, United States}
\affil[2]{School of Systems and Enterprises, Stevens Institute of Technology, Hoboken, New Jersey, 07030, United States}


\date{September, 2024}
\maketitle

\begin{abstract}
Understanding the complexity and interdependence of systems in the Anthropocene is essential for making informed decisions about societal challenges spanning geophysical, biophysical, sociocultural, and sociotechnical domains. This paper explores the potential of Hetero-functional Graph Theory (HFGT) as a quantification tool for converting Model-based Systems Engineering (MBSE), stated in the Systems Modeling Language (SysML), into dynamic simulations—offering a comprehensive alternative to System Dynamics (SD) for representing interdependent systems of systems in the Anthropocene. The two approaches are compared in terms of systems thinking abstractions, methodological flexibility, and their ability to represent dynamic, multi-functional systems. Through a comparative study, the Mono Lake system is simulated in Northern California using both SD and MBSE \& HFGT to highlight technical, conceptual and analytical differences. The simulations show equivalent results.  However, MBSE \& HFGT provide distinct advantages in capturing the nuances of the system through a broader set of systems thinking abstractions and in managing adaptive, multi-functional system interactions. These strengths position MBSE \& HFGT as a powerful and flexible approach for representing, modeling, analyzing, and simulating heterogeneous and complex systems-of-systems in the Anthropocene.
 
\end{abstract}

\section{Introduction}
Human activities have induced profound and irreversible changes to Earth systems, leading to numerous, complex, and intertwined societal challenges of the Anthropocene \cite{little2023earth,verburg2016methods,steffen2018trajectories}.   Here, the Anthropocene refers to a period where humans have become dominant agents of change in Earth system dynamics\cite{zalasiewicz2020anthropocene,liu2007coupled,steffen2007anthropocene,goudie2018human,lewis2018human} with Anthropocene systems including geophysical, biophysical, sociocultural and sociotechnical systems \cite{little2023earth}.  Human-induced stresses include rapid population growth, widespread urbanization, and an increase in the exploitation of natural resources \cite{mondal2022challenges}.  These alter the trajectory of Earth systems and cause deepening social, technical and ecological challenges \cite{kotze2019earth,mazac2020post}.  The interconnected nature of these challenges arises from their shared underlying systems where perturbations in one system cascade through others with complex interactions\cite{folke2021our,williams2015anthropocene,malhi2017concept}. For example, deforestation reduces biodiversity by destroying habitats, exacerbates climate change by eliminating carbon sinks, and affects agriculture by destabilizing local hydrological cycles.  Anthropocene challenges are numerous and wide-ranging, including global warming, over-exploitation of natural resources, freshwater scarcity, habitat loss, and environmental pollution that not only threatens many species with extinction, but also jeopardizes the health of human society\cite{steffen2011anthropocene}.  Such challenges extend across nearly every discipline including economics, public health, ecology, food, energy and natural resources.  Unfortunately, however, they are usually tackled in isolation, with only a few recent attempts to develop comprehensive approaches that recognize their intertwined nature\cite{little2023earth, an2014agent,steffen2011anthropocene,reyers2018social}. Although some studies recognize the interconnected nature of challenges arising from human-induced alterations in Anthropocene systems \cite{xiaoming2018linking,li2023challenges,liu2007coupled}, researchers are often unable to capture the synergies and trade offs across multiple systems due to the lack of understanding \cite{berkes2008navigating,robinson2018modelling} or availability of analytical frameworks to quantify these intertwined relationships\cite{little2023earth, li2023challenges}. 

Recognizing the interdependent nature of societal challenges of the Anthropocene, researchers have sought to study and ultimately coordinate the underlying system-of-systems.   This includes coupled human-natural systems \cite{liu2007coupled,hamilton2015integrated,liu2007complexity,an2012modeling}, integrated assessment models\cite{halog2011advancing,weyant2017some}, water-land-energy-carbon nexus\cite{zhao2018impacts,feng2024impact,jiang2023spatio}, food-energy-water (FEW) nexus \cite{naderi2021system, putra2020systematic, zhang2018water,wang2023system}, and its extensions to the carbon cycle\cite{chamas2021sustainable,yadav2021food,xu2020impacts},  the environment\cite{correa2022novel,yue2021managing,mperejekumana2024integrating}, and the climate\cite{radini2021urban,schmidt2018state,afkhami2024sustainability}. Indeed, the widespread demand for integrating and coordinating various interconnected systems to tackle complex societal challenges is clear\cite{little2023earth}. When analyzing complex systems to identify the trade-offs and synergies, a wide range of modeling approaches has been used. These include system thinking\cite{arnold2015definition,cabrera2023systems} , system dynamics (SD)\cite{elsawah2017overview,berrio2021understating,francisco2023food}, multi-layer networks \cite{milanovic2017modeling,kenett2014network}, Bayesian networks \cite{shi2020coupling,chai2020quantifying}, agent-based modeling \cite{an2014agent,monticino2007coupled,kline2017integrating}, and optimization\cite{zhang2020synergy,li2019stochastic}. Such human-natural systems exhibit different levels of complexity \cite{moallemi2020exploratory,li2023challenges}.  They are inherently heterogeneous, with different components, structures, processes, interactions, and scales \cite{an2012modeling}.  Consequently, existing modeling approaches often fail to address the increasing complexity caused by the interdependence of these systems\cite{ostrom2009general,norberg2008complexity}.

Although these studies have used a plethora of modeling and analysis approaches\cite{wu2023social,han2023assessing,martinez2021integral,dentoni2021linking,gotts2019agent,karan2018towards}, SD is frequently adopted \cite{martinez2021integral,berrio2021understating,oliveira2022socio,wen2022system}.  Pioneered in the late 1950s, SD is a modeling approach designed to study, simulate, and analyze systems in various application domains including: water resources\cite{mirchi2012synthesis,winz2009use},  economics\cite{radzicki2020system}, socio-ecology\cite{berrio2021understating,elsawah2017overview}, renewable energy\cite{laimon2022system,fontes2018sustainable}, the food industry\cite{minegishi2000system} and healthcare\cite{faezipour2013system}.  Users of SD often report a user-friendly environment for rapidly building, calibrating, and testing mathematical models composed of multiple differential equations \cite{sterman2002system,ogata2004system}.   Consequently, it is widely adopted as a modeling approach that captures dynamic interactions within human-natural systems and helps decision-makers anticipate the impacts of different policies \cite{azar2012system}.  Rooted in system thinking and control theory\cite{sterman2002system}, SD employs causal loop diagrams (CLDs) to capture feedback mechanisms between system parameters.  It also uses stock and flow diagrams (SFDs) to organize the model into stock and flow variables.  These two diagrams (which are often integrated) serve to model the time-evolution of systems with continuous, often non-linear, dynamics with multiple interacting elements\cite{madani2009system}.  Furthermore, the graphical nature of CLDs and SFDs facilitate their use in participatory processes that aid in teaching, building trust and collective decision-making\cite{stave2010participatory,voinov2018tools}.  In all, SD has demonstrated the ability to manage several types of system complexity found in human-natural systems including non-linearity, feedback mechanisms, and potential instability\cite{elsawah2017overview}.  

While the use of SD has certainly advanced our knowledge of human-natural systems, the SD approach utilizes a fairly limited set of systems thinking abstractions.  As elaborated below, Anthropocene systems-of-systems exhibit many additional complexities that require a broader set of system thinking abstractions to be understood and managed.  Consequently, a modeling and analysis approach that utilizes only a few systems thinking abstractions may make rapid and early advances due to its ease of use, but is likely to encounter methodological limitations that impede its ability to address the remaining complexities found in these Anthropocene systems-of-systems.

\subsection{Original Contribution}
This paper proposes model-based systems engineering (MBSE) and hetero-functional graph theory (HFGT) as an alternative modeling and analysis approach to SD. MBSE \& HFGT collectively 
\begin{enumerate*}
\item exhibit a broader set of systems thinking abstractions, 
\item reproduce the analytical conclusions produced with an SD approach, and \item ultimately overcome many of the methodological limitations that may arise in Anthropocene systems-of-systems.
\end{enumerate*}
More specifically, this paper introduces SD in terms of its constituent systems thinking abstractions.   It elaborates how these abstractions have served to address many of the intertwined societal challenges of the Anthropocene, and identifies where methodological limitations have been encountered.  It then introduces MBSE \& HFGT in terms of its constituent systems thinking abstractions.  This introduction highlights where this broader set of systems thinking abstractions has direct relevance to the complexities found in Anthropocene systems-of-systems.  The paper then introduces HFGT as a means of developing a computational model from a graphical model stated in the Systems Modeling Language (SysML).   This MBSE \& HFGT approach is demonstrated on Mono Lake (California) as a hydrological system that has been previously studied with system dynamics.  This demonstration confirms that MBSE \& HFGT can reproduce the modeling and simulation results developed through an SD approach. Thus, it demonstrates practically that no analytical insight is lost by adopting MBSE \& HFGT as an alternative to SD.  To the contrary, this practical demonstration highlights how MBSE \& HFGT can provide greater analytical insight -- even for the same example system.  Finally, the paper revisits the methodological limitations encountered with SD and proposes how they may be addressed with MBSE \& HFGT. 

\subsection{Paper Outline}
The remainder of the paper is organized as follows.  Sec. \ref{Sec:SDModeling} provides an assessment of SD in terms of its constituent systems thinking abstractions. Sec. \ref{Sec:MBSEHFGT} provides a similar assessment for MBSE \& HFGT. Sec. \ref{Sec:HFGT} provides essential background definitions on HFGT and Sec. \ref{Sec:Monoillustrativeexample} demonstrates the application of MBSE \& HFGT to Mono Lake as a hydrological system.  Sec. \ref{Sec:DiscussionandResults} then discusses the implications of these results on the potential for MBSE \& HFGT to address the interdependent societal challenges of the Anthropocene.  Sec. \ref{Sec:Conclusion} brings the paper to a conclusion.

\section{Systems Thinking Assessment of System Dynamics}\label{Sec:SDModeling}
This section provides an assessment of SD in terms of its constituent systems thinking abstractions.   While a comprehensive exposition of SD cannot be provided here, a brief overview of SD is provided in Sec. \ref{Sec:SDOverview}.  Next, Sec. \ref{Sec:SDAssessment} provides an overview of the systems thinking assessment in terms of three main types of systems thinking abstractions:  system boundary, system form, and system function.  Each of these are then elaborated in Sec. \ref{Sec:SDSystemBoundary}, \ref{Sec:SDSystemForm}, and \ref{Sec:SDSystemFunction} respectively.  
\liinesbigfig{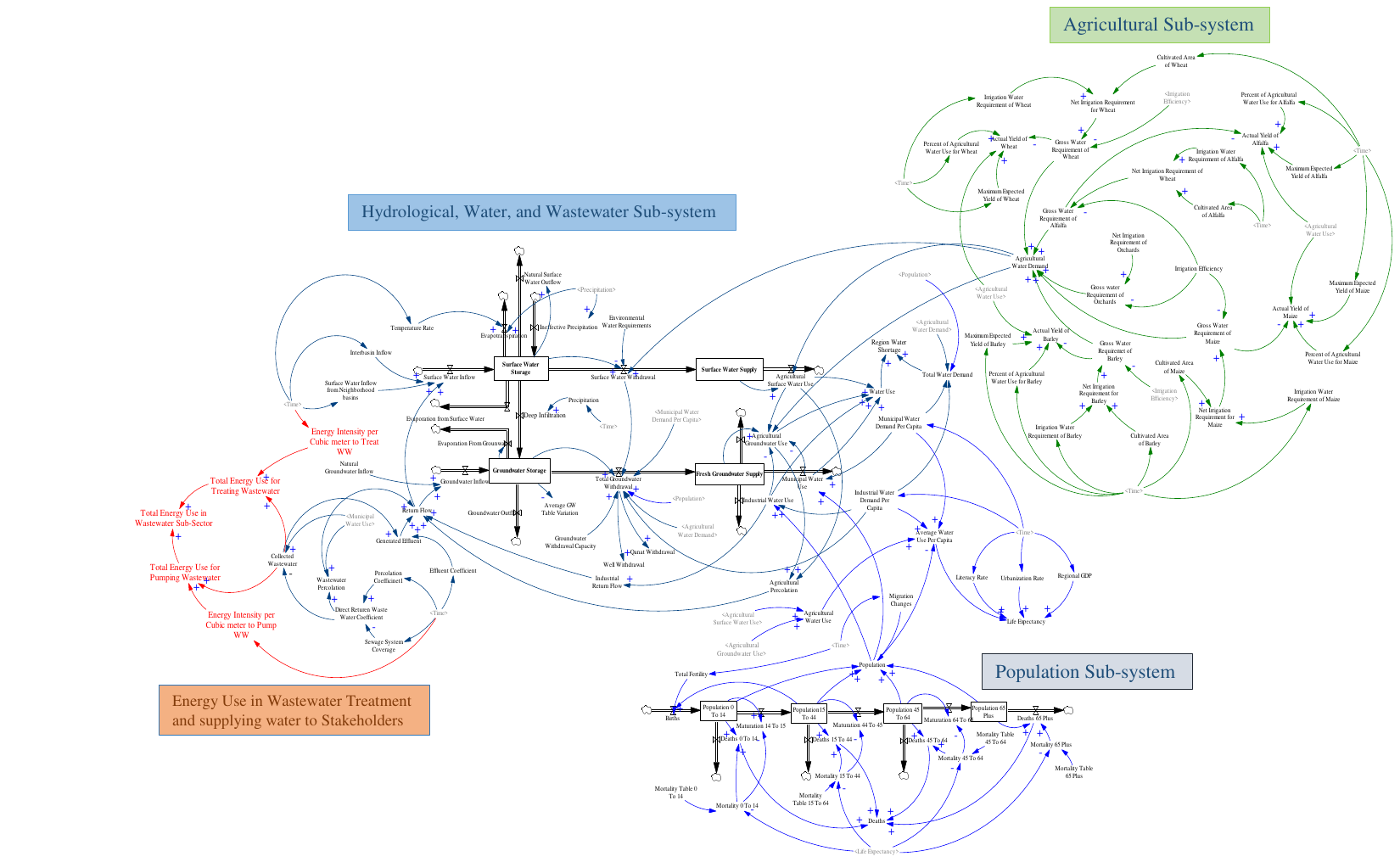}{An integrated SD SFD and CLD of the food-energy-water system of Qazvin Plain in Iran\cite{naderi2021system}.}{Fig:FEW-SD}{6.5}
\subsection{Overview of System Dynamics}\label{Sec:SDOverview}
Fig. \ref{Fig:FEW-SD} can be used to illustrate the SD approach with an SFD of a FEW system in the Qazvin plain in Iran\cite{naderi2021system} as a representative Anthropocene system-of-systems. The complex FEW system has four sub-systems: 
\begin{enumerate*}
\item agricultural, 
\item population, 
\item energy, and 
\item hydrological and water resources.
\end{enumerate*}
The model provides a high-level representation that includes the production of major crops, water supply and demand, the urban wastewater system, population, and energy use. Dynamic, non-linear feedback loops are represented with (peer-to-peer) arrows between variables in the system.  The directionality of the arrows indicates the (assumed) direction of causality and the polarity of the arrows defines the sign of the causality\cite{richardson1995loop}.  For example, Fig. \ref{Fig:FEW-SD} denotes the positive effect of precipitation on deep infiltration with a positive polarity arrow while it denotes the negative effect of municipal water use on agricultural groundwater use with a negative polarity arrow.  Note that Fig. \ref{Fig:FEW-SD} does not depict a system boundary because a core assumption of the SD approach is that the \emph{system model} is closed\cite{cosenz2016applying,wolstenholme2003towards,sterman2002system,richardson1997introduction,forrester1997industrial}.  Nevertheless, SD can model an open system by distinguishing between endogenous variables and exogenous parameters (depicted in text).  Endogenous variables (e.g., total water demand, return flow, sewage system coverage) take on time-varying values that can be affected by the execution of the model while exogenous parameters (e.g., temperature, surface water inflow from neighboring basins and precipitation) take on time-varying input values that cannot be affected by the execution of the model.  These exogenous parameters are assumed to originate from outside the system despite being inside the system model.  An SD model also has constant endogenous parameters (e.g., lake capacity, river width) tied to the structure of the system.  Additionally, SD employs two types of core components:
\begin{enumerate*}
\item stocks that represent the accumulation/depletion of objects, and 
\item flows that denote their time rate of change as an output signal\cite{stave2003system}.
\end{enumerate*}
The linkages between these stocks and flows describes the dynamics of how the objects are consumed, replenished, or transformed.  The SD approach assumes that the system evolves with continuous (i.e., non-discrete), continuous-time processes, and is best suited when these continuous processes exhibit extensive feedback mechanisms.  System dynamic models are often simulated to evaluate policies and explore ``what-if" scenarios \cite{ahmad2004spatial,zomorodian2018state,madani2009system,naderi2021system,richardson1995loop}.  Time-varying exogenous parameters as well as constant endogenous parameters can be adjusted to create multiple computational experiments that describe how endogenous system variables respond.   Consequently,  these computational experiments allow policymakers and stakeholders to simulate the impacts of proposed interventions, identify potential unintended consequences, and compare alternative strategies for achieving desired outcomes\cite{stave2003system}.

\subsection{Overview of Systems Thinking Assessment}\label{Sec:SDAssessment}
The \emph{formal} study of systems thinking asserts that all systems can be understood in terms of generalized systems thinking abstractions.  Here, these abstractions are understood to capture the fundamental or intrinsic nature of an entity; focusing on what is most important or universally true about it while neglecting its concrete realities in terms of specific objects or instances\cite{dictionary}.  In systems engineering and science, systems thinking abstractions are the primary means by which scientists manage the \emph{cognitive} complexity associated with the \emph{intrinsic} complexity of systems\cite{crawley2015system}.  As the scientist expands their palette of abstractions in their systems thinking, they are able to reduce their \emph{perception} of the cognitive complexity of the system, although the \emph{intrinsic} complexity of the system remains unchanged.  Although systems thinking abstractions can be and are often discussed entirely abstractly, it is often pedagogically helpful to provide specific examples that exhibit these abstractions as an initial means of understanding the abstractions more generically.  Here, again, the system dynamic model presented in Fig. \ref{Fig:FEW-SD} serves as a useful example by which to discuss the abstractions and assess the degree to which SD modeling exhibits them.  While not every system exhibits all systems thinking abstractions, all systems axiomatically\cite{crawley2015system,suh1998axiomatic} exhibit three broad categories of systems thinking abstractions:  
\begin{enumerate*}
\item Systems Boundary,
\item System Form, and 
\item System Function.
\end{enumerate*}
The presence of each of these abstractions in systems dynamics modeling is now discussed, with a summary provided in Table \ref{ta:SD-HFGTabstractions}.  

\begin{table}[htpb!]
\caption{Strengths and Limitations of SD compared to MBSE-SysML}\vspace{-0.1in}
\label{ta:SD-HFGTabstractions}
\includegraphics[width=\linewidth]{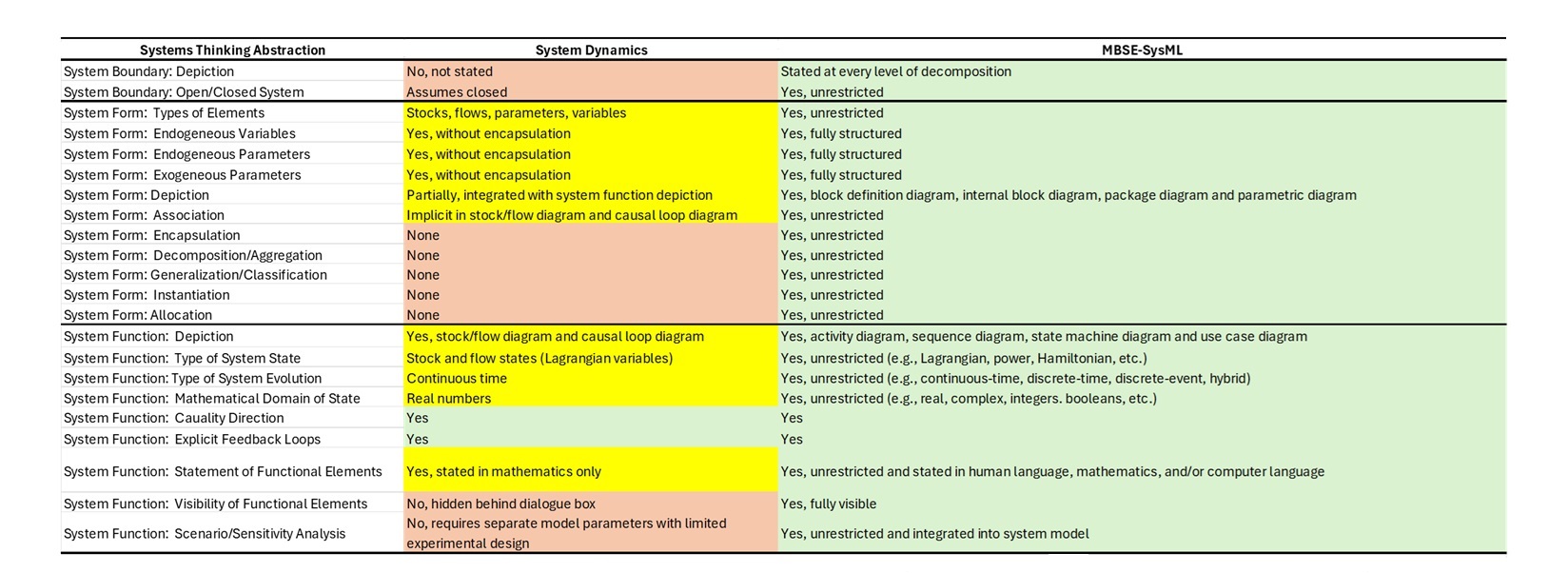}
\end{table}

\subsection{Assessing System Boundary}\label{Sec:SDSystemBoundary}

The system boundary delineates between the system itself and everything else (in the system context).  In many graphical modeling approaches, the system boundary is often drawn as a closed shape (e.g., rectangle or oval).  When artifacts are allowed to pass through the system boundary (to the system as inputs or from the system as outputs), the system is said to be an open system. If not, it is a closed system.  As shown in Fig. \ref{Fig:FEW-SD}, in SD, the system boundary is not \emph{depicted} explicitly, although it very much still exists.  The system boundary associated with the FEW system in Fig. \ref{Fig:FEW-SD} may be thought of as the figure boundary or the application window in which the system's computational model resides.  The absence of a depicted system boundary in SD modeling forces an assumption of a closed system which brings about an obfuscation.  For example, the FEW system in Fig. \ref{Fig:FEW-SD} has ``interbasin inflow" as an exogenous time-varying parameter.  Such parameters are often thought of as system inputs that cross the system boundary.  However, in SD modeling, such an exogenous time-varying parameter is understood as an internal flow that has been mathematically predetermined (for simplifying the model or reducing complexity).  It is effectively indistinguishable in type from, for example, an internal heartbeat.  Additionally, the naming of the parameter as an ``inflow" when it is already inside the system creates another ambiguity tied to the question: ``Into where (but another system)?"  Furthermore, the absence of an explicitly depicted system boundary precludes the possibility of first-pass, by inspection, analyses such as matter/energy conservation.  While any open system model can be converted into a mathematically equivalent closed system, the absence of a \emph{depicted} system boundary obfuscates whether system elements are truly part of the system or not, or why parameters may be exogenous or endogenous.  This obfuscation adds to the cognitive complexity of the system and detracts from the meaning that can be inferred from the SD model.  While the ``interbasin inflow" example was a simple one, the inputs and outputs of a (large and complex) Anthropocene system-of-systems are likely to be much less clear.  Equally importantly, the inputs and outputs \emph{between} systems in the system-of-systems are also not demarcated in system-dynamics modeling.  Moreover, many well-known abstract concepts in human society, such as scope-of-contract, scope-of-work and jurisdiction, cannot be graphically represented until a system boundary is explicitly depicted.  In the context of the FEW system, SD lacks the capability to explicitly represent the decision-making responsibilities of government agencies embedded within the agricultural, energy, hydrological, and wastewater subsystems, or the regulatory frameworks established by authorities to govern each sector. Last but not least, the absence of an explicit boundary could further impede our ability to understand the system's form and function.  

\subsection{Assessing System Form}\label{Sec:SDSystemForm}
The system form describes the elements that make up the system, their constituent parts and attributes, and how they are connected and/or related. The ability of system dynamics to generically model system form is summarized in Table \ref{ta:SD-HFGTabstractions}.  SD uses three types of meta-elements or modeling primitives:  stocks, flows and parameters (related to stocks and flows).  The choice of these modeling primitives aligns with a paradigm (or mental model) entrenched in stocks and flows.  Consequently, when a modeler encounters an application domain (e.g., business or the environment) where the prevailing paradigm is based on stocks and flows, they find that SD modeling can be straightforwardly applied.  While this advantage yields rapid and early advances, it is important to recognize that many systems do not have natural stock and flow representations and instead use completely different sets of modeling primitives\cite{farid2016engineering}.  For example, GIS-data use points, lines, and polygons as modeling primitives\cite{congalton1992abcs}.  Networked systems use graphs composed of nodes and edges\cite{myers2003software}.  Physical systems use generalized capacitors, inductors, resistors, transformers, and gyrators\cite{tellegen1948gyrator}.  Consequently, SD modeling is likely to encounter limitations when Anthropocene systems-of-systems require a broader set of meta-elements (or modeling primitives).  

Returning to Table \ref{ta:SD-HFGTabstractions}, SD models include many attributes that can describe endogenous variables, endogenous parameters, and exogenous parameters.  Again, as stated above, the distinction between these in SD is not clear due to the absence of a system boundary.  Furthermore, in most cases, these attributes appear ``stand-alone" on the SFDs and CLDs without any mention of how they are encapsulated within larger elements.  For example, Fig. \ref{Fig:FEW-SD} shows many parameters tied to wastewater treatment plants without encapsulating them within a larger wastewater treatment plant element.  Thus, while SD shows the causal relationships between these attributes, it does not reveal how they are encapsulated together. Consequently, as the heterogeneity and size of the SD model grows, the CLDs and SFDs often grow into a cumbersome and flat depiction of highly entangled low-level parameters.  Indeed, the use of CLDs/SFDs is limiting because it confounds system form and system function.  Normally, causal \emph{interactions} (i.e., the interactions between the system's functions) are part of the system function while peer-to-peer associations show the interfaces between elements of the system form.  Functional interactions and formal interfaces (i.e., associations) are only equivalent when the allocation of function to form is one-to-one\cite{crawley2015system}.  

In addition to not depicting any form of encapsulation, these diagrams do not show decomposition or aggregation relationships where certain elements are depicted as parts of other elements.  In Anthropocene systems-of-systems, part-whole relationships are a natural extension of the system boundary abstraction.  SD diagrams also do not show generalization or classification relationships.  In Anthropocene systems-of-systems, such relationships are often needed to distinguish between different types of people (e.g., ethnicity), land (e.g., agricultural, suburban, urban), or water bodies (e.g., lakes, rivers, reservoirs).  SD diagrams also do not show instantiation relationships.  In many systems, it is often useful to characterize the relationship between elements generically rather than specifying these relationships for each and every instance.  For example, in watersheds, river segments generically flow downstream to intersections which flow into river segments again.  This generic pattern constitutes a \emph{reference architecture}\cite{crawley2015system,farid2015axiomatic} which can serve to greatly reduce the cognitive complexity associated with our understanding of these systems.  In contrast, SD forces the modeler to graphically specify each and every stock, flow, and parameter.  Beyond the graphical clutter this creates, it becomes increasingly tedious to specify each and every stock and flow instance manually.  Consequently, many SD models (as shown in Fig. \ref{Fig:FEW-SD}) tend to limit the total number of stocks and flows in the system to several handful.  In contrast, automated model development approaches can easily include millions of stocks, flows, and their associated parameters.  Finally, SD models do not address the allocation of system function to elements of system form.  In Anthropocene systems-of-systems, many parts of the system carry out more than one function.  For example, a hospital can carry out many different types of procedures.  

\subsection{Assessing System Function}\label{Sec:SDSystemFunction}

The system function describes what the system does, how it behaves, and for human systems, its reason for existence \cite{crawley2015system}. The ability of SD to generically model system function is summarized in Table \ref{ta:SD-HFGTabstractions}.   

Again, the choice of stocks and flows as modeling primitives in SD affects its approach to understanding a system's function. From the perspective of system function, a stock stores some type of quantity (e.g., matter, energy, information, etc.) while a flow transports a type of quantity. Consequently, while a stock and flow diagram depicts system function for systems with these modeling primitives, it does not depict the function of systems that require different modeling primitives.  Additionally, the choice of stocks and flows as modeling primitives forces a choice of system state with Lagrangian variables\cite{bertsekas2014constrained} -- where the associated state parameters are related to each other by a derivative of time.  In many other system models, the system state can be described by power variables or Hamiltonian variables\cite{leimkuhler2004simulating}.   While the choice of Lagrangian, power, or Hamiltonian variables is not particularly important in linear dynamic systems, it can have a significant impact on the the tractability of non-linear dynamic systems; particularly if they are large and complex.  System dynamic models also assume an evolution based upon continuous-time differential equations (even if the underlying simulation software solves these differential equations discretely).   In contrast, many Anthropocene systems may evolve over discrete-time or have discrete-event behavior.  System dynamic models also assume that state variables are real-valued.   In contrast, many Anthropocene systems-of-systems may have state variables that are defined over integers or complex numbers.  Despite these limitations in system dynamics modeling, SFDs as well as CLDs are able to show the direction of causality and explicit feedback loops.  Both of these concepts are essential to understanding the associated complexities in a system's function.  

The practice of SD modeling is often intrinsically tied to SD modeling software (e.g., Vensim or Stella).  These software tools have their own limitations.   While the relationship between model attributes can be described as a mathematical function, it may be useful to relate parameters in natural human language or in a computer programming language.  The former lends itself to participatory modeling with a diversity of (non-mathematical) stakeholders, while the latter lends itself to complex relationships beyond a (relatively simple) algebraic expression.  It is also worth noting that these functional relationships are often hidden within dialogue boxes and thus can only be viewed one at a time.   Finally, many of these software tools have a limited ability to conduct scenario/sensitivity analyses.  While it is possible to define/design several scenarios, these tools have not been developed to investigate extensive uncertainty in model inputs via robust experimental designs (e.g., large-scale Monte Carlo simulations).   

In conclusion, this discussion shows that an SD approach utilizes a fairly limited set of systems thinking abstractions.  Consequently, using a SD approach is likely to impose methodological limitations that impede its ability to address many of the complexities found in Anthropocene systems-of-systems.   
 
\section{Systems Thinking Assessment of Model-Based Systems Engineering \& the Systems Modeling Language}
\label{Sec:MBSEHFGT}
Given the limited set of systems thinking abstractions found in an SD approach, this paper proposes MBSE and SysML as an alternative.  More specifically, this section discusses how MBSE when implemented using the Systems Modeling Language admits a broader set of systems thinking abstractions.  Again, although systems thinking abstractions can be and are often discussed entirely abstractly, it may be pedagogically helpful to provide specific examples that exhibit these abstractions as an initial means of understanding these abstractions more generically.  Now, in this case, the HFGT meta-architecture in Fig. \ref{fig:HFGT_BDD} and \ref{fig:HFGT_ACT} serves as a useful example by which to discuss the abstractions and assess the degree to which SysML exhibits them.  More specifically, Fig. \ref{fig:HFGT_BDD} shows the Block Definition Diagram (BDD) of the HFGT meta-architecture while Fig. \ref{fig:HFGT_ACT} shows its corresponding ACTivity Diagram (ACT).  The former is primarily focused on representing system form, while the latter is primarily focused on representing system function.  Unlike SD where a single graphical model is depicted, SysML enables multiple diagrams that provide complementary and mutually consistent ``views" of the system.  As is often emphasized: \emph{``A diagram of the model is never the model itself, it is merely one view of the model."}\cite{delligatti2013sysml}.  Different types of diagrams (e.g., BDDs and ACTs) serve different modeling purposes.  Additionally, any given diagram need not necessarily include every element of the system, as long as the diagrams collectively and unambiguously represent the system.  This strategy facilitates the representation of large complex systems without too much graphical clutter; thereby managing the cognitive load imposed on the human reader of the graphical model.  As done previously, the remainder of the discussion proceeds along the three broad categories of systems thinking abstractions:  
\begin{enumerate*}
\item Systems Boundary,
\item System Form, and 
\item System Function
\end{enumerate*}
with a summary provided in Table \ref{ta:SD-HFGTabstractions}.  

\liinesbigfig{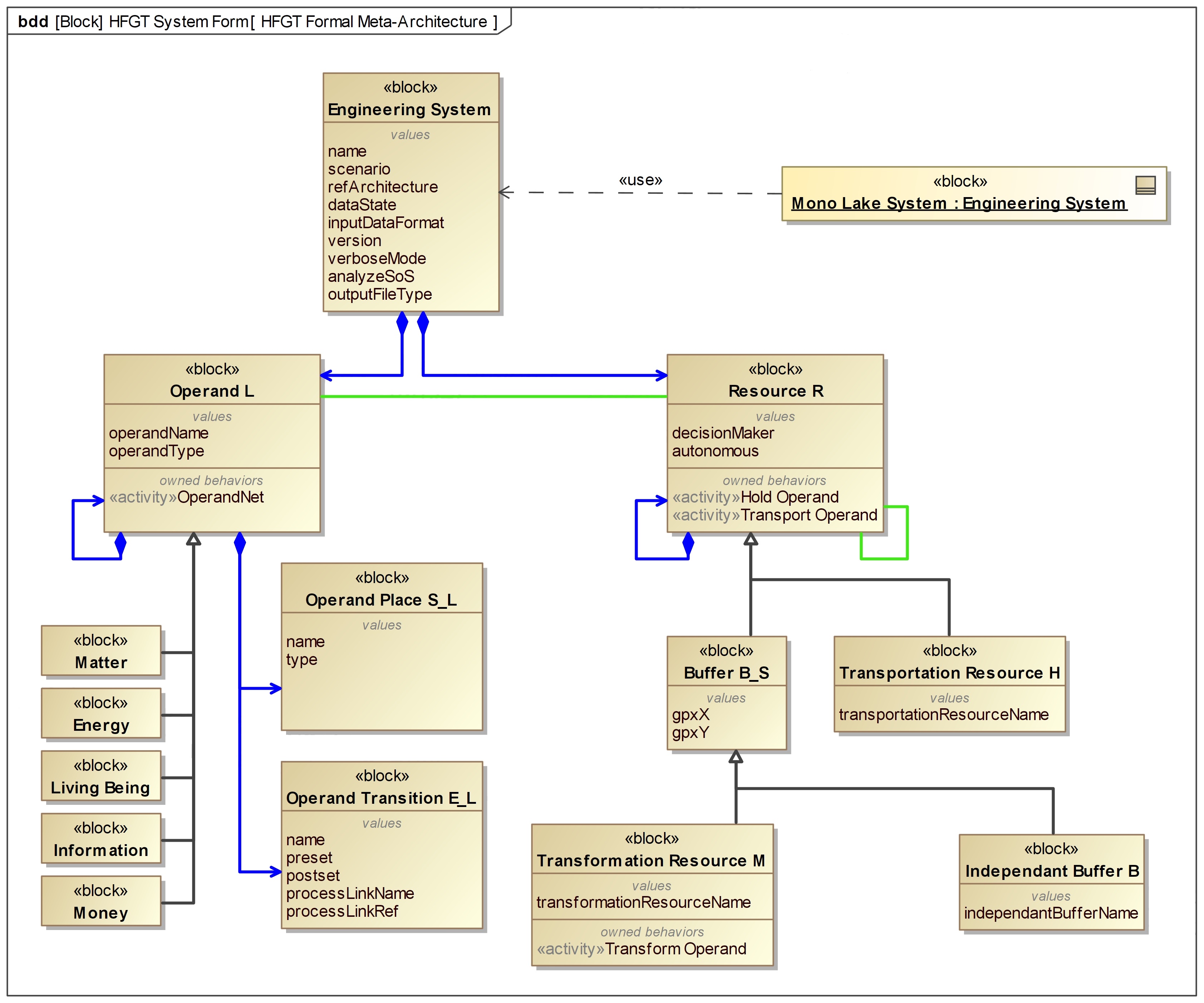}{A SysML BDD Diagram: the meta-architecture of the system form (adapted from \cite{schoonenberg2019hetero}).}{fig:HFGT_BDD}{5}

\liinesbigfig{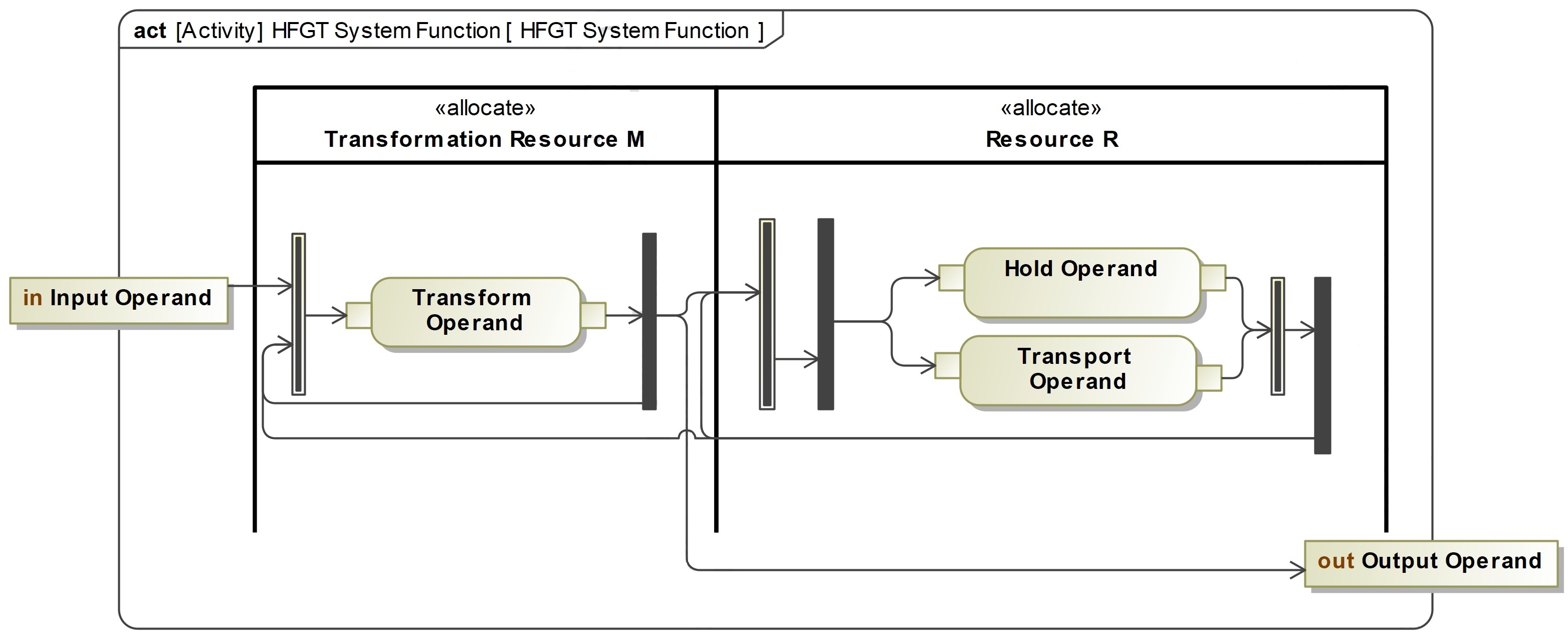}{A SysML ACT Diagram: the meta-architecture of the system function (adapted from \cite{schoonenberg2019hetero}).}{fig:HFGT_ACT}{6}

\subsection{System Boundary}
In SysML, the system boundary is always explicitly depicted with the boundary of the diagram regardless of which type of diagram is being used.  Furthermore, this system boundary is labelled on the top left with four pieces of information in order: 
\begin{enumerate*}
\item the type of diagram (e.g., BDD, ACT, etc.), 
\item the type of model element (e.g., a block, activity, etc.), 
\item the name of the model element, and
\item the name of the diagram.      
\end{enumerate*}
All four of these are required. The model element name is required because in reality a system requires multiple system boundaries; one for the system itself and one for each of its elements -- be they component blocks, subsystems, or even activities.  The diagram name is required because more than one diagram may be required to depict a named model element.  The type of diagram and model element are required so that the graphical syntax of the diagram is correct and consistent with the remainder of the model.  Consequently, Fig. \ref{fig:HFGT_BDD} shows ``bdd" as the type of diagram, ``[Model]" as the type of model element, ``HFGT System Form" as the name of the model element and ``HFGT Formal Meta-Architecture" as the name of the diagram.  Similarly, Fig. \ref{fig:HFGT_ACT} shows ``act" as the type of diagram, ``[Activity]" as the type of model element, ``HFGT System Function" as the name of the model element, and ``[HFGT Functional Meta-Architecture]" as the name of the diagram.  In addition to the diagram boundary (of a model element), the boundary of a given model element serves as the system boundary of that model element when that model element is considered as a system.  

In SysML, the open vs. closed nature of the system boundary is always explicitly depicted.  For example, in Fig. \ref{fig:HFGT_BDD} all of the blocks that have a connecting arrow are open systems in their own right.  Similarly, in Fig. \ref{fig:HFGT_ACT} all activities that have a connecting arrow are open system activities in their own right.  The absence of such connecting arrows would represent a closed system.  Finally, Fig. \ref{fig:HFGT_ACT} itself depicts an activity as an open system because it shows an input parameter called ``Input Operand" and an output parameter called ``Output Operand".  The absence of such input and output parameters would represent a closed system.    

\subsection{System Form}
As shown in Table \ref{ta:SD-HFGTabstractions}, SysML fully supports the representation of system form.  Four types of diagrams can be used to represent the system form.  
\begin{itemize}
\item The Block Definition Diagram (BDD) primarily represents the hierarchy of the system form.
\item The Internal Block Diagram (IBD) primarily represents the internal structure of the instantiated system.  
\item The Package Diagram (PKG) organizes model elements into model packages (that are not necessarily aligned with the hierarchy of the system).  
\item The Parametric Diagram (PAR) is a special type of IBD that relates model parameters to each other.  
\end{itemize}
Of the four diagrams, only the BDD (e.g., Fig. \ref{fig:HFGT_BDD}) is formally required when modeling a system while the other three provide supporting context-specific information.  More specifically, a BDD contains blocks that represent the system's constituent elements.  Unlike SD where only stocks, flows, and parameters are allowed as modeling primitives, a BDD places no restrictions on the nature of its blocks and its constituent parameters.  In this way, SysML can include any set of modeling primitives that a modeler wishes to use.  Perhaps more importantly in the context of Anthropocene systems-of-systems, multiple sets of modeling primitives can be combined and reconciled into a single SysML model.  A BDD also supports endogenous variables, endogenous parameters, and exogenous parameters without restriction as attributes that pertain to blocks.  The question of whether a given attribute is endogenous, exogenous, a parameter or a variable does not pertain to the system form, and is instead represented in a diagram that pertains to the systems behavior (e.g., an ACT).  Unlike in SD where attributes are loosely depicted in CLDs, in SysML all attributes are encapsulated within blocks and consequently are organized with the hierarchy of the system's form.  

In addition to the above, SysML fully supports the systems thinking abstractions tied to the relationships between elements of form.  The first systems thinking abstraction is peer-to-peer association where two elements transfer or exchange quantities of matter, energy, information, money, or living organisms. SysML depicts associations with a solid line between two blocks.  For example, in Fig. \ref{fig:HFGT_BDD}, there is an association link between  Operand $(L)$ and Resources $(R)$ blocks, meaning that a connection can exist between instances of those blocks in the operational system. The second systems thinking abstraction is decomposition or aggregation where a system is broken into smaller pieces or constituents of the form.  In other words, assembling individual constituents into a unified whole is referred to as aggregation, whereas breaking down a system into its constituent parts is known as decomposition. SysML depicts decomposition or aggregation with a solid line between two blocks with a solid diamond on one end. An open arrowhead on the part end of the line conveys unidirectional access from the composite to its part; the absence of an arrowhead conveys bidirectional access.  For example, Fig. \ref{fig:HFGT_BDD}, shows that Engineering System is composed of Operand $(L)$ and Resources $(R)$.   The third systems thinking abstraction is a classification or generalization where one element is a type of another.  This relationship conveys inheritance between two elements: a more generalized element, called the supertype, and a more specialized element, the subtype. Generalizations are used to create classification trees (type hierarchies) in system model.  SysML depicts classification with  a solid line with a hollow, triangular arrowhead on the end of the supertype.  For example, in Fig. \ref{fig:HFGT_BDD}, Buffer $(B_S)$ and Transportation Resource $(H)$ are two blocks generalized as Resources $(R)$. Or, it can also be read as Resource $(R)$ classifies into Buffer $(B_S)$ and Transportation Resource $(H)$. The fourth systems thinking abstraction is instantiation where an element appears as an instance of another.  SysML depicts instantiation with a block whose name includes the name of the instance followed by a colon (:) followed by the name of the class-block being instantiated.  For example, Fig. \ref{fig:HFGT_BDD} shows that Mono Lake system is an instantiation of Engineering System.  Although SysML (Version 1.X) does not (currently) have a dedicated type of arrow relation for instantiation as a systems thinking abstraction, an arrow has been depicted in Fig. \ref{fig:HFGT_BDD} to emphasize the dependence of Mono Lake system's instance on the engineering system block in the HFGT meta-architecture.  The fifth systems thinking abstraction is allocation where an element of function (i.e., an activity) is allocated to an element of form (i.e., a block).  SysML depicts allocation with an activity serving as a component behavior of a block.  For example, in Fig. \ref{fig:HFGT_BDD}, Transport Operand is the behavior allocated to the Resources $(R)$.

\subsection{System Function}
As shown in Table \ref{ta:SD-HFGTabstractions}, SysML fully supports the representation of system function with four types of diagrams: 
\begin{itemize}
\item The ACTivity Diagram (ACT) primarily represents the parallel and serial arrangement of functions, processes or activities in the system function. 
\item The Sequence Diagram (SEQ) primarily represents the functional interactions between blocks as sequences of operand exchanges. 
\item The State Machine Diagram (STM) describes the state and transitions of a given block in the system.  
\item The Use Case Diagram (UC) describes how external users called actors interact with the system function.  
\end{itemize}
Of the four diagrams, the ACT (e.g., Fig. \ref{fig:HFGT_ACT}) is the most commonly used and is effectively required for most non-trivial systems.  The other three diagrams provide supporting context-specific information.

Together, these four diagrams fully support the systems thinking abstractions identified in Table \ref{ta:SD-HFGTabstractions}. Fig. \ref{fig:HFGT_ACT}, shows the ACT diagram of the HFGT meta-architecture which serves as a clear example to explain the system thinking abstractions in terms of system function. First, unlike system dynamics, SysML places no restriction on the type of system state.  For example, Lagrangian (e.g., stock and flow), power, and Hamiltonian variables can be defined in a BDD over real numbers, complex numbers, integers, or even booleans and then subsequently evolved in an ACT.  Consequently, SysML straightforwardly admits continuous-time, discrete-time, discrete-event, and hybrid state evolutions. In Fig. \ref{fig:HFGT_ACT}, vertical swimlanes correspond to Resources $(R)$  and Transformation Resource $(M)$ blocks of the HFGT meta-architecture. The owned activities of these two blocks are shown within the swimlanes as rounded rectangles. Variables embedded within each of these three activities of the system can be defined with different system state and evolution. Also, it should be noted that the arrow-head lines in Fig. \ref{fig:HFGT_ACT} show the object flows of matter, energy, information, etc. throughout the system. The state evolutions can be represented in an ACT while the SEQ and STM diagrams provide additional functionality for discrete-event behavior.  All three of these diagrams (e.g., Fig. \ref{fig:HFGT_ACT}) show the direction of causality and include an explicit depiction of feedback loops.   All three diagrams can state these functional elements in natural human language, in mathematical language, or in a computer language.  Such flexibility assists in the analysis and synthesis of complex systems at various levels of abstraction and technical detail.   Finally, scenario/sensitivity analyses can be parameterized within a BDD and then described functionally with either an ACT or STM.  Overall, SysML's rich graphical ontology fully supports the representation of the systems thinking abstractions tied to system function.  

In conclusion, this discussion shows that MBSE and SysML utilize a rich set of systems thinking abstractions with full support for a detailed understanding of system boundary, system form, and system function.  Consequently, using MBSE and SysML is likely to overcome the methodological limitations that may be encountered with an SD approach.  Furthermore, this broad support for systems thinking abstractions is likely to facilitate our understanding of the complexities found in Anthropocene systems-of-systems.

\section{Fundamentals of Hetero-Functional Graph Theory}\label{Sec:HFGT}
While MBSE, and more specifically SysML, provide a common modeling language\cite{little2023earth} for large, complex systems-of-systems, it does not have built-in functionality for conducting quantitative analysis.   Fortunately, HFGT\cite{Schoonenberg:2019:ISC-BK04,Farid:2022:ISC-J51,Farid:2016:ISC-BC06} provides an analytical means of translating graphical SysML models into mathematical and computational models.  HFGT has been applied to numerous application domains individually, including electric power\cite{Farid:2015:SPG-J17,Thompson:2021:SPG-J46}, potable water\cite{Farid:2015:ISC-J19}, transportation\cite{Viswanath:2013:ETS-J08}, and mass-customized production systems\cite{Farid:2015:IEM-J23,Farid:2008:IEM-J05,Farid:2008:IEM-J04,Farid:2017:IEM-J13}.  Perhaps more importantly, HFGT has already demonstrated its applicability to systems-of-systems including multi-modal electrified transportation systems\cite{Farid:2016:ETS-J27,vanderWardt:2017:ETS-J33,Farid:2016:ETS-BC05}, microgrid-enabled production systems\cite{Schoonenberg:2017:IEM-J34}, personalized healthcare delivery systems\cite{Khayal:2015:ISC-J20,Khayal:2017:ISC-J35,Khayal:2021:ISC-J48}, hydrogen-natural gas systems\cite{Schoonenberg:2022:ISC-J50}, energy-water-nexus\cite{Farid:2024:ISC-JR04}, and the American multi-modal energy system\cite{Thompson:2024:ISC-J55}.  Interestingly, and in contrast to SD models, several of these systems-of-systems applications of HFGT combine both time-driven as well as discrete-event phenomena.

\subsection{Essential Definitions}\label{subsec:HFGT}
Hetero-functional graphs (HFGs) are formal graphical models that represent the interconnectedness of complex systems-of-systems. HFGT is conceptually rooted in the universal structural principles of human language in terms of subjects and predicates, where predicates consist of verbs and objects\cite{Schoonenberg:2019:ISC-BK04,Farid:2022:ISC-J51}.  This reliance on the universal structure of human-language provides the basis for a discipline-agnostic ontology that facilitates cross-disciplinary applications. 
 More specifically, HFGT translates the meta-architecture shown in Figs. \ref {fig:HFGT_BDD} and \ref{fig:HFGT_ACT}.  A system is defined by a set of resources $R$, which act as the subjects; a set of processes $P$, which serve as predicates; and a set of operands $L$, which constitute the objects involved in these processes.   
\begin{defn}[System Operand 
\cite{SE-Handbook-Working-Group:2015:00}]
\label{Defn:D1}
An asset or object $l_i \in L$ that is operated on or utilized in the course of a process execution.
\end{defn}
\begin{defn}[System Process
\cite{Hoyle:1998:00,SE-Handbook-Working-Group:2015:00}]
\label{def:CH7:process}
An activity $p \in P$ that converts or transfers a specified set of input operands into a designated set of outputs. 
\end{defn}
\begin{defn}[System Resource
\cite{SE-Handbook-Working-Group:2015:00}]
An asset or object $r_v \in R$ that enables the execution of a process.  
\end{defn}
Returning to Fig. \ref{fig:HFGT_BDD}, the system resources $R=M \cup B \cup H$ are categorized into transformation resources $M$, independent buffers $B$, and transportation resources $H$.  The set of ``buffers" $B_S=M \cup B$ is introduced to support the discussion, and the system processes $P = P_\mu \cup P_{\bar{\eta}}$ are classified into transformation processes $P_\mu$ and refined transportation processes $P_\eta$.  This occurs from the concurrent execution of a transportation process and a holding process. HFGT further highlights that resources can perform one or more system processes, thereby generating a set of ``capabilities"\cite{Schoonenberg:2019:ISC-BK04}.
\begin{defn}[Buffer
\cite{Schoonenberg:2019:ISC-BK04,Farid:2022:ISC-J51}]
\label{defn:BSCh7}

A resource $r_v \in R$ is a buffer $b_s \in B_S$  if and only if it has the capacity to store or transform one or more operands at a specific spatial location.  
\end{defn}
\begin{defn}[Capability\cite{Schoonenberg:2019:ISC-BK04,Farid:2022:ISC-J51,Farid:2016:ISC-BC06}]
\label{defn:capabilityCh7}
An action $e_{wv} \in {\cal E}_S$ (in the SysML sense) defined by a system process $p_w \in P$ being executed by a resource $r_v \in R$.  It constitutes a subject + verb + operand sentence of the form: ``Resource $r_v$ does process $p_w$".  
\end{defn}

The engineering system meta-architecture found in Figs. \ref{fig:HFGT_BDD} and \ref{fig:HFGT_ACT} must be instantiated and eventually translated into corresponding Petri net model(s) \cite{Farid:2022:ISC-J51}. To facilitate this, the positive and negative hetero-functional incidence tensors are introduced to characterize the flow of operands through buffers and capabilities.  
\begin{defn}[The Negative 3$^{rd}$ Order Hetero-functional Incidence Tensor (HFIT) $\widetilde{\cal M}_\rho^-$
\cite{Farid:2022:ISC-J51}]
\label{Defn:D6}
The negative hetero-functional incidence tensor $\widetilde{\cal M_\rho}^- \in \{0,1\}^{|L|\times |B_S| \times |{\cal E}_S|}$  is a third-order tensor whose element $\widetilde{\cal M}_\rho^{-}(i,y,\psi)=1$ when the system capability ${\epsilon}_\psi \in {\cal E}_S$ pulls operand $l_i \in L$ from buffer $b_{s_y} \in B_S$.
\end{defn} 

\begin{defn}[The Positive  3$^{rd}$ Order Hetero-functional Incidence Tensor (HFIT) $\widetilde{\cal M}_\rho^+$
\cite{Farid:2022:ISC-J51}]
The positive hetero-functional incidence tensor $\widetilde{\cal M}_\rho^+ \in \{0,1\}^{|L|\times |B_S| \times |{\cal E}_S|}$  is a third-order tensor whose element $\widetilde{\cal M}_\rho^{+}(i,y,\psi)=1$ when the system capability ${\epsilon}_\psi \in {\cal E}_S$ injects operand $l_i \in L$ into buffer $b_{s_y} \in B_S$.
\end{defn}
\noindent These incidence tensors are straightforwardly ``matricized" to form second-order Hetero-functional Incidence Matrices $M = M^+ - M^-$ with dimensions $|L||B_S|\times |{\cal E}|$. Consequently, the supply, demand, transportation, storage, transformation, assembly, and disassembly of multiple operands in distinct locations over time can be described by an Engineering System Net and its associated State Transition Function\cite{Schoonenberg:2022:ISC-J50}.

\begin{defn}[Engineering System Net
\cite{Schoonenberg:2022:ISC-J50}]
\label{Defn:ESN}
An elementary Petri net ${\cal N} = \{S, {\cal E}_S, \textbf{M}, W, Q\}$, where:
\begin{itemize}
    \item $S$ is the set of places with size: $|L||B_S|$,
    \item ${\cal E}_S$ is the set of transitions with size: $|{\cal E}|$,
    \item $\textbf{M}$ is the set of arcs, with the associated incidence matrices: $M = M^+ - M^-$,
    \item $W$ is the set of weights on the arcs, as captured in the incidence matrices,
    \item $Q=[Q_B; Q_E]$ is the marking vector for both the set of places and the set of transitions. 
\end{itemize}
\end{defn}

\begin{defn}[Engineering System Net State Transition Function
\cite{Schoonenberg:2022:ISC-J50}]
\label{Defn:ESN-STF}
The  state transition function of the engineering system net $\Phi()$ is:
\begin{equation}\label{CH6:eq:PhiCPN}
Q[k+1]=\Phi(Q[k],U^-[k], U^+[k]) \quad \forall k \in \{1, \dots, K\}
\end{equation}
where $k$ is the discrete time index, $K$ is the simulation horizon, $Q=[Q_{B}; Q_{\cal E}]$, $Q_B$ has size $|L||B_S| \times 1$, $Q_{\cal E}$ has size $|{\cal E}_S|\times 1$, the input firing vector $U^-[k]$ has size $|{\cal E}_S|\times 1$, and the output firing vector $U^+[k]$ has size $|{\cal E}_S|\times 1$.  
\begin{align}\label{CH6:CH6:eq:Q_B:HFNMCFprogram}
Q_{B}[k+1]&=Q_{B}[k]+{M}^+U^+[k]\Delta T-{M}^-U^-[k]\Delta T \\ \label{CH6:CH6:eq:Q_E:HFNMCFprogram}
Q_{\cal E}[k+1]&=Q_{\cal E}[k]-U^+[k]\Delta T +U^-[k]\Delta T
\end{align}
where $\Delta T$ is the duration of the simulation time step.  
\end{defn}
\noindent It is crucial to mention that the engineering system net’s state transition function explicitly embodies continuity laws, allowing both the Eulerian and Lagrangian perspectives depending on the required modeling application.

In addition to the engineering system net, in HFGT, each operand can have its own state and evolution.  This behavior is described by an Operand Net and its associated State Transition Function for each operand. 
\begin{defn}[Operand Net\cite{Farid:2008:IEM-J04,Schoonenberg:2019:ISC-BK04,Khayal:2017:ISC-J35,Schoonenberg:2017:IEM-J34}]\label{Defn:OperandNet} Given operand $l_i$, an elementary Petri net ${\cal N}_{l_i}= \{S_{l_i}, {\cal E}_{l_i}, \textbf{M}_{l_i}, W_{l_i}, Q_{l_i}\}$ where: 
\begin{itemize}
\item $S_{l_i}$ is the set of places describing the operand's state.  
\item ${\cal E}_{l_i}$ is the set of transitions describing the evolution of the operand's state.
\item $\textbf{M}_{l_i} \subseteq (S_{l_i} \times {\cal E}_{l_i}) \cup ({\cal E}_{l_i} \times S_{l_i})$ is the set of arcs, with the associated incidence matrices: $M_{l_i} = M^+_{l_i} - M^-_{l_i} \quad \forall l_i \in L$.  
\item $W_{l_i} : \textbf{M}_{l_i}$ is the set of weights on the arcs, as captured in the incidence matrices $M^+_{l_i},M^-_{l_i} \quad \forall l_i \in L$.  
\item $Q_{l_i}= [Q_{Sl_i}; Q_{{\cal E}l_i}]$ is the marking vector for both the set of places and the set of transitions. 
\end{itemize}
\end{defn}

\begin{defn}[Operand Net State Transition Function\cite{Farid:2008:IEM-J04,Schoonenberg:2019:ISC-BK04,Khayal:2017:ISC-J35,Schoonenberg:2017:IEM-J34}]\label{Defn:OperandNet-STF}
The  state transition function of each operand net $\Phi_{l_i}()$ is:
\begin{equation}\label{CH6:eq:PhiSPN}
Q_{l_i}[k+1]=\Phi_{l_i}(Q_{l_i}[k],U_{l_i}^-[k], U_{l_i}^+[k]) \quad \forall k \in \{1, \dots, K\} \quad i \in \{1, \dots, L\}
\end{equation}
where $Q_{l_i}=[Q_{Sl_i}; Q_{{\cal E} l_i}]$, $Q_{Sl_i}$ has size $|S_{l_i}| \times 1$, $Q_{{\cal E} l_i}$ has size $|{\cal E}_{l_i}| \times 1$, the input firing vector $U_{l_i}^-[k]$ has size $|{\cal E}_{l_i}|\times 1$, and the output firing vector $U^+[k]$ has size $|{\cal E}_{l_i}|\times 1$.  

\begin{align}\label{X}
Q_{Sl_i}[k+1]&=Q_{Sl_i}[k]+{M_{l_i}}^+U_{l_i}^+[k]\Delta T - {M_{l_i}}^-U_{l_i}^-[k]\Delta T \\ \label{CH6:CH eq:Q_E:HFNMCFprogram}
Q_{{\cal E} l_i}[k+1]&=Q_{{\cal E} l_i}[k]-U_{l_i}^+[k]\Delta T +U_{l_i}^-[k]\Delta T
\end{align}
\end{defn}

\vspace{0.5cm}
\subsection{The Hetero-functional Network Minimum Cost Flow (HFNMCF) Problem}

HFGT simulates the behavior of an engineering system using the Hetero-Functional Network Minimum Cost Flow (HFNMCF) problem\cite{Schoonenberg:2022:ISC-J50}.  The HFNMCF problem extends the classical network minimum cost flow problem to account for the heterogeneity of function found in Anthropocene systems-of-systems.  It optimizes the time-dependent flow and storage of multiple operands between buffers, allows for their transformation from one operand to another, and tracks the state of these operands.  In this regard, it is a very flexible optimization problem that applies to a wide variety of complex engineering systems.  For the purposes of this paper, the HFNMCF problem is a type of discrete-time-dependent, time-invariant, convex optimization program\cite{Schoonenberg:2022:ISC-J50}.

\begin{align}\label{Eq:ObjFunc}
Z &= \sum_{k=1}^{K-1} x^T[k] F_{QP} x[k] + f_{QP}^T x[k]
\end{align}
\begin{align}\label{Eq:ESN-STF1}
\text{s.t. } -Q_{B}[k+1]+Q_{B}[k]+{M}^+U^+[k]\Delta T - {M}^-U^-[k]\Delta T=&0 && \!\!\!\!\!\!\!\!\!\!\!\!\!\!\!\!\!\!\!\!\!\!\!\!\!\!\!\!\!\!\!\!\!\!\!\!\!\!\!\!\!\forall k \in \{1, \dots, K\}\\  \label{Eq:ESN-STF2}
-Q_{\cal E}[k+1]+Q_{\cal E}[k]-U^+[k]\Delta T + U^-[k]\Delta T=&0 && \!\!\!\!\!\!\!\!\!\!\!\!\!\!\!\!\!\!\!\!\!\!\!\!\!\!\!\!\!\!\!\!\!\!\!\!\!\!\!\!\!\forall k \in \{1, \dots, K\}\\ \label{Eq:DurationConstraint}
 - U_\psi^+[k+k_{d\psi}]+ U_{\psi}^-[k] = &0 && \!\!\!\!\!\!\!\!\!\!\!\!\!\!\!\!\!\!\!\!\!\!\!\!\!\!\!\!\!\!\!\!\!\!\!\!\!\!\!\!\!\forall k\in \{1, \dots, K\} \quad \psi \in \{1, \dots, {\cal E}_S\}\\ \label{Eq:OperandNet-STF1}-Q_{Sl_i}[k+1]+Q_{Sl_i}[k]+{M}_{l_i}^+U_{l_i}^+[k]\Delta T - {M}_{l_i}^-U_{l_i}^-[k]\Delta T=&0 && \!\!\!\!\!\!\!\!\!\!\!\!\!\!\!\!\!\!\!\!\!\!\!\!\!\!\!\!\!\!\!\!\!\!\!\!\!\!\!\!\!\forall k \in \{1, \dots, K\} \quad i \in \{1, \dots, |L|\}\\ \label{Eq:OperandNet-STF2}
-Q_{{\cal E}l_i}[k+1]+Q_{{\cal E}l_i}[k]-U_{l_i}^+[k]\Delta T + U_{l_i}^-[k]\Delta T=&0 && \!\!\!\!\!\!\!\!\!\!\!\!\!\!\!\!\!\!\!\!\!\!\!\!\!\!\!\!\!\!\!\!\!\!\!\!\!\!\!\!\!\forall k \in \{1, \dots, K\} \quad i \in \{1, \dots, |L|\}\\ \label{Eq:OperandNetDurationConstraint}
- U_{xl_i}^+[k+k_{dxl_i}]+ U_{xl_i}^-[k] = &0 &&  \!\!\!\!\!\!\!\!\!\!\!\!\!\!\!\!\!\!\!\!\!\!\!\!\!\!\!\!\!\!\!\!\!\!\!\!\!\!\!\!\!
\forall k\in \{1, \dots, K\}, \: \forall x\in \{1, \dots, |{\cal E}_{l_i}\}|, \: l_i \in \{1, \dots, |L|\}\\ \label{Eq:SyncPlus}
U^+_L[k] - \widehat{\Lambda}^+ U^+[k] =&0 && \!\!\!\!\!\!\!\!\!\!\!\!\!\!\!\!\!\!\!\!\!\!\!\!\!\!\!\!\!\!\!\!\!\!\!\!\!\!\!\!\!\forall k \in \{1, \dots, K\}\\ \label{Eq:SyncMinus}
U^-_L[k] - \widehat{\Lambda}^- U^-[k] =&0 && \!\!\!\!\!\!\!\!\!\!\!\!\!\!\!\!\!\!\!\!\!\!\!\!\!\!\!\!\!\!\!\!\!\!\!\!\!\!\!\!\!\forall k \in \{1, \dots, K\}\\ \label{CH6:eq:HFGTprog:comp:Bound}
\begin{bmatrix}
D_{Up} & \mathbf{0} \\ \mathbf{0} & D_{Un}
\end{bmatrix} \begin{bmatrix}
U^+ \\ U^-
\end{bmatrix}[k] =& \begin{bmatrix}
C_{Up} \\ C_{Un}
\end{bmatrix}[k] && \!\!\!\!\!\!\!\!\!\!\!\!\!\!\!\!\!\!\!\!\!\!\!\!\!\!\!\!\!\!\!\!\!\!\!\!\!\!\!\!\!\forall k \in \{1, \dots, K\} \\\label{Eq:OperandRequirements}
\begin{bmatrix}
E_{Lp} & \mathbf{0} \\ \mathbf{0} & E_{Ln}
\end{bmatrix} \begin{bmatrix}
U^+_{l_i} \\ U^-_{l_i}
\end{bmatrix}[k] =& \begin{bmatrix}
F_{Lpi} \\ F_{Lni}
\end{bmatrix}[k] && \!\!\!\!\!\!\!\!\!\!\!\!\!\!\!\!\!\!\!\!\!\!\!\!\!\!\!\!\!\!\!\!\!\!\!\!\!\!\!\!\!\forall k \in \{1, \dots, K\}\quad i \in \{1, \dots, |L|\} \\\label{CH6:eq:HFGTprog:comp:Init} 
\begin{bmatrix} Q_B ; Q_{\cal E} ; Q_{SL} \end{bmatrix}[1] =& \begin{bmatrix} C_{B1} ; C_{{\cal E}1} ; C_{{SL}1} \end{bmatrix} \\ \label{CH6:eq:HFGTprog:comp:Fini}
\begin{bmatrix} Q_B ; Q_{\cal E} ; Q_{SL} ; U^- ; U_L^- \end{bmatrix}[K+1] =   &\begin{bmatrix} C_{BK} ; C_{{\cal E}K} ; C_{{SL}K} ; \mathbf{0} ; \mathbf{0} \end{bmatrix}\\ \label{ch6:eq:QPcanonicalform:3}
\underline{E}_{CP} \leq D(X) \leq& \overline{E}_{CP} \\ \label{Eq:DeviceModels}
g(X,Y) =& 0 \\ \label{Eq:DeviceModels2}
h(Y) \leq& 0 
\end{align}
where $X=\left[x[1]; \ldots; x[K]\right]$  is the vector of primary decision variables, and $Y=\left[y[1]; \ldots; y[K]\right]$  is the vector of auxiliary decision variables at time $k$.

\subsubsection{Objective Function}
In Eq.  \ref{Eq:ObjFunc}, $Z$ is a convex objective function separable in discrete time steps $k$.  Matrix $F_{QP}$ and vector $f_{QP}$ allow quadratic and linear costs to be incurred from the place and transition markings in both the engineering system net and operand nets:
\begin{itemize}
\item $F_{QP}$ is a positive semi-definite, diagonal, quadratic coefficient matrix.
\item $f_{QP}$ is a linear coefficient matrix.
\end{itemize}

\subsubsection{Equality Constraints}

\begin{itemize}
\item Eqs.  \ref{Eq:ESN-STF1} and \ref{Eq:ESN-STF2} describe the state transition function of an engineering system net (Defn. \ref{Defn:ESN} \& \ref{Defn:ESN-STF}).
\item Eq \ref{Eq:DurationConstraint} is the engineering system net transition duration constraint where the end of the $\psi^{th}$ transition occurs $k_{d\psi}$ time steps after its beginning. 
\item Eqs.  \ref{Eq:OperandNet-STF1} and \ref{Eq:OperandNet-STF2} describe the state transition function of each operand net ${\cal N}_{l_i}$ (Defn. \ref{Defn:OperandNet} \& \ref{Defn:OperandNet-STF}) associated with each operand $l_i \in L$.  
\item Eq.  \ref{Eq:OperandNetDurationConstraint} is the operand net transition duration constraint where the end of the $x^{th}$ transition occurs $k_{dx_{l_i}}$ time steps after its beginning. 
\item Eqs. \ref{Eq:SyncPlus} and \ref{Eq:SyncMinus} are synchronization constraints that couple the input and output firing vectors of the engineering system net to the input and output firing vectors of the operand nets, respectively. $U_L^-$ and $U_L^+$ are the vertical concatenations of the input and output firing vectors $U_{l_i}^-$ and $U_{l_i}^+$, respectively.
\begin{align}
U_L^-[k]&=\left[U^-_{l_1}; \ldots; U^-_{l_{|L|}}\right][k] \\
U_L^+[k]&=\left[U^+_{l_1}; \ldots; U^+_{l_{|L|}}\right][k]
\end{align}
\item Eqs.  \ref{CH6:eq:HFGTprog:comp:Bound} and \ref{Eq:OperandRequirements} are boundary conditions.  Eq. \ref{CH6:eq:HFGTprog:comp:Bound} is a boundary condition constraint that allows some of the engineering system net firing vector decision variables to be set to an exogenous constant.  Eq. \ref{Eq:OperandRequirements} does the same for the operand net firing vectors.  
\item Eqs.  \ref{CH6:eq:HFGTprog:comp:Init} and \ref{CH6:eq:HFGTprog:comp:Fini} are the initial and final conditions of the engineering system net and the operand nets where $Q_{SL}$ is the vertical concatenation of the place marking vectors of the operand nets $Q_{Sl_i}$.
\begin{align}
Q_{SL}^-[k]&=\left[Q^-_{Sl_1}; \ldots; U^-_{Sl_{|L|}}\right][k] \\
U_{SL}^+[k]&=\left[U^+_{Sl_1}; \ldots; U^+_{Sl_{|L|}}\right][k]
\end{align}
\end{itemize}

\subsubsection{Inequality Constraints}
 $D_{QP}()$ and vector $E_{QP}$ in Eq.  \ref{ch6:eq:QPcanonicalform:3} place capacity constraints on the vector of primary decision variables at each time step $x[k] = \begin{bmatrix} Q_B ; Q_{\cal E} ; Q_{SL} ; Q_{{\cal E}L} ; U^- ; U^+ ; U^-_L ; U^+_L \end{bmatrix}[k] \quad \forall k \in \{1, \dots, K\}$. 

\vspace{0.1in}
\subsubsection{Device Model Constraints}
g(X,Y) and h(Y) in Eqs. \ref{Eq:DeviceModels} and \ref{Eq:DeviceModels2} respectively are a set of device model functions whose presence and nature depend on the specific problem application.  They can not be further elaborated until the application domain and its associated capabilities are identified.  

\section{Illustrative Example:  Mono Lake system}\label{Sec:Monoillustrativeexample}
The relatively abstract discussions of SD, MBSE, and HFGT in Sec. \ref{Sec:SDModeling}, \ref{Sec:MBSEHFGT}, and \ref{Sec:HFGT} respectively can now be applied to a practical and illustrative example.  More specifically, this section demonstrates the application of MBSE \& HFGT to Mono Lake as a hydrological system that has been previously studied with system dynamics.  Subsection \ref{Sec:MonoLakeOverview} provides an overview of Mono Lake.  Subsection \ref{Sec:SDMonoLake} provides a graphical description of an SD model of the Mono Lake system.  Subsection \ref{mathematicalequationsofmono} then elaborates this SD model mathematically.  Given this formal description of the Mono Lake system, Subsection \ref{MonoSysML} describes Mono Lake graphically in SysML relative to the HFGT meta-architecture.  Finally, Subsection \ref{HFNMCFmathematics} translates the SD mathematical model to the HFNMCF problem.   

\liinesbigfig{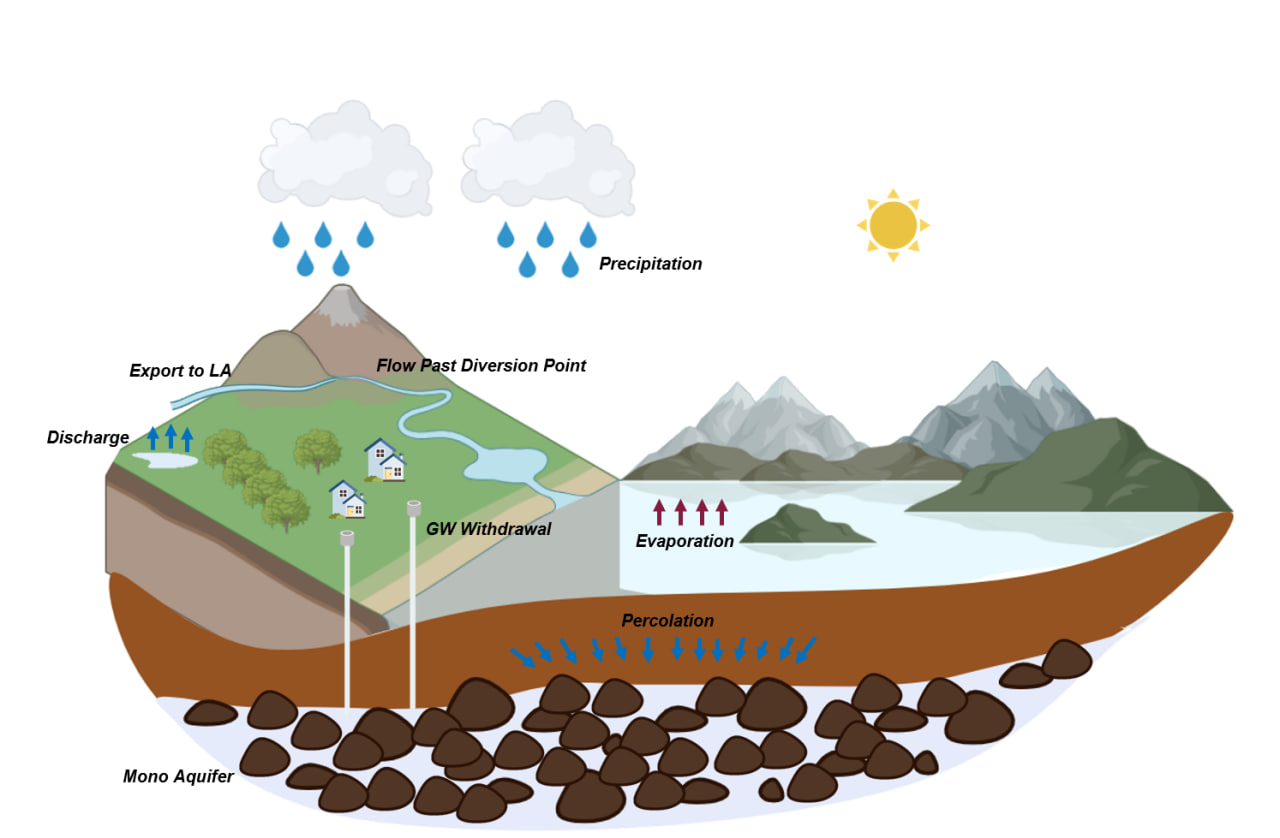}{Conceptual hydrological and water resources model of Mono Lake in California.}{fig:monoconceptualfigure}{7}
\subsection{Overview of Mono Lake, CA}\label{Sec:MonoLakeOverview}
Located in the eastern Sierra Nevada Mountains of California, USA (Fig. \ref{fig:monoconceptualfigure}), Mono Lake is a hyper-saline, closed basin primarily receiving water from surrounding streams, with loss of water primarily by evaporation \cite{phillips2021microbial}. The region has an arid climate with an average precipitation of about 1.2 feet per year. Beneath Mono Lake lies the Mono Aquifer, supplying about 6800 kilo acre feet (KAF) of water, mainly for agricultural purposes.  Before entering the lake, a considerable portion of water used to be diverted to Los Angeles, causing a significant drop in the water level. However, in 1994, the California State Water Resources Control Board stopped water diversion to reduce the demand on the lake as a water resource. 

\subsection{Graphical Description of System Dynamics Model of Mono Lake}\label{Sec:SDMonoLake}
The SD model of Mono Lake \cite{ford2010modeling} is developed using the Vensim software \cite{Vetana}. The SD model visually represents the lake's variables and their relationships with each other as a causal loop diagram (Fig. \ref{fig:CLD&SFD}), and computationally simulates the system's behavior based on a stock flow diagram (Fig. \ref{fig:CLD&SFD}). As previously mentioned, CLDs have polarized arrows to show a positive (same direction) or negative (opposite direction) change among the components of the system, which may form balancing or reinforcing feedback loops \cite{mirchi2012synthesis,naderi2021system,sterman2002system}. The SD model reveals how various components and their relationships influence the lake dynamics, providing a tool to simulate the lake's behavior under different scenarios \cite{sterman2002system, naderi2021system}.
\liinesbigfig{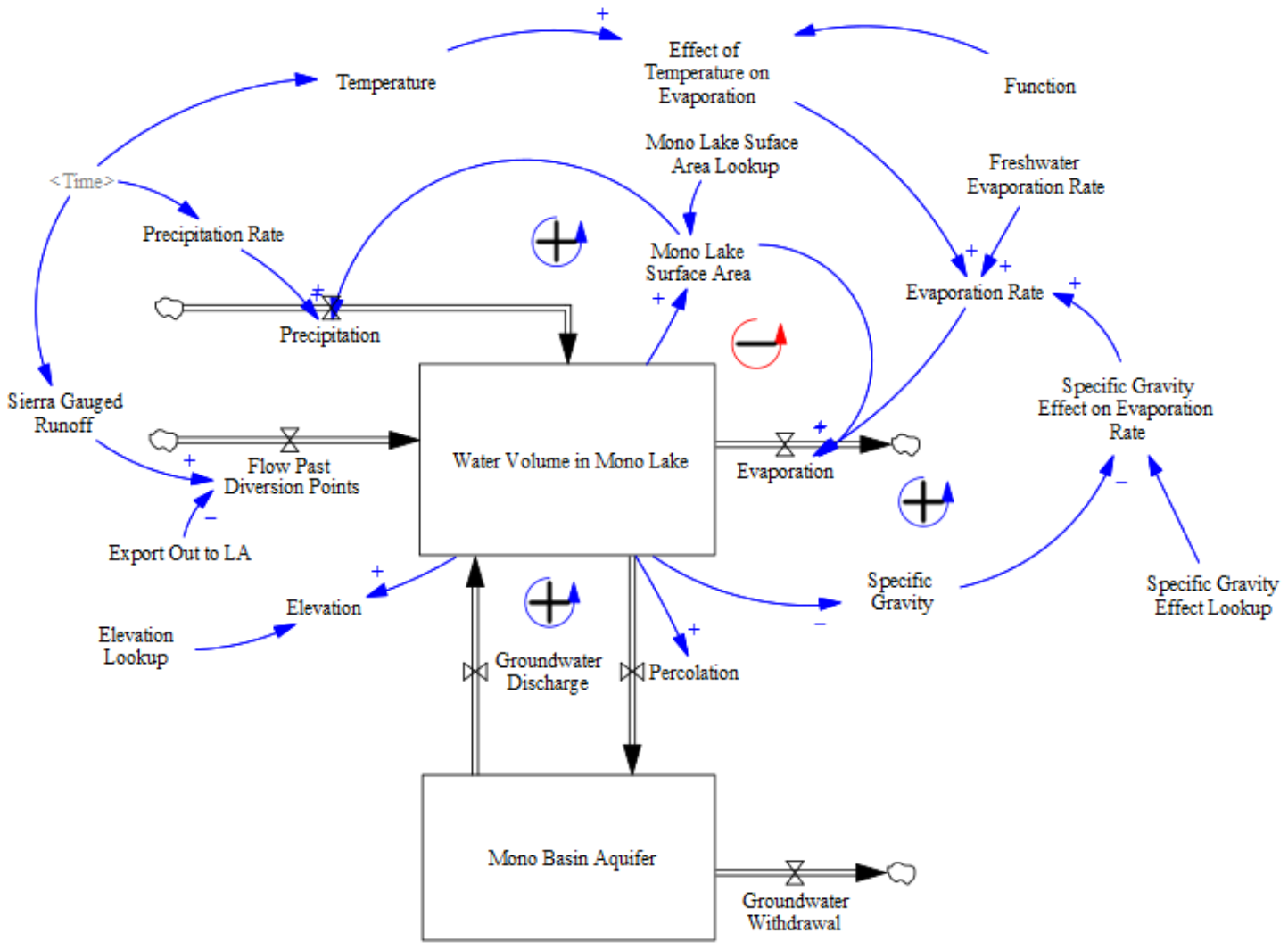}{Integrated SD CLD and SFD of Mono Lake system.}{fig:CLD&SFD}{6.5}

Fig. \ref{fig:CLD&SFD} shows the interactions between the hydrologic processes in the Mono Lake system.  Temperature, precipitation rate and Sierra Gauged Runoff are exogenously defined and are outside the system boundary. Although these variables never directly change in response to feedback, they impact the endogenous variables within the system. The lake receives volumetric flow rates including:  water as precipitation, the  flow past diversion point, and the groundwater discharge.  It also loses water through evaporation and percolation (Fig. \ref{fig:CLD&SFD}).  The aquifer receives water via percolation from the lake, with water loss through natural discharge and withdrawal. As illustrated in Fig. \ref{fig:CLD&SFD}, three reinforcing loops and one balancing loop control the system behavior. Lookup tables are used to model the elevation ($\text{H}[k]$) and surface area (${A}_{\text{S}}$) of the lake as it varies with the lake's water volume.  Similarly, the effects of temperature and specific gravity on evaporation rate are formulated using linear lookup functions.

\subsection{Mathematical Description of System Dynamics Model of Mono Lake} \label{mathematicalequationsofmono}
This section describes the mathematics underlying the system dynamics model of the Mono Lake system in Eqs. \ref{Eq:waterco} to \ref{eq:effect of specific gravity}.

\begin{align}
V_{\text{Mono}}[k+1] - V_{\text{Mono}}[k] - \left(\dot{V}_{\text{P}}[k] + \dot{V}_{\text{FPDP}}[k] + \dot{V}_{\text{Dis}}[k] - \dot{V}_{\text{Evap}}[k] - \dot{V}_{\text{Perc}}[k]\right) \Delta T &= 0 && \forall k \in \{0, \ldots, K\} \label{Eq:waterco}\\
V_{\text{Aqui}}[K+1] - V_{\text{Aqui}}[k] - \left(\dot{V}_{\text{Perc}}[k] + \dot{V}_{\text{Dis}}[k] + \dot{V}_{\text{GWWith}}[k]\right) \Delta T &= 0 && \forall k \in \{0, \ldots, K\} \label{Eq:watercontinuitylawforaquifer} \\
\dot{V}_{\text{GDis}}[k] &= 
\dot{V}^{\ddagger}_{\text{GDis}}
&& \forall k \in \{0, \ldots, K\}  \label{eq:groundwaterdischarge}\\
\dot{V}_{\text{GWWith}}[k] &= \dot{V}_{\text{GWWith}}^\ddagger  && \forall k \in \{0, \ldots, K\} \label{eq:constantwithdrawl} \\
V_{\text{Mono}}[1] &= V^{\ddagger}_{0-\text{Mono}} && \forall k \in \{0, \ldots, K\}\label{eq:initialwatervolumeinlake}\\
V_{\text{Aqui}}[1] &= V^{\ddagger}_{0-\text{Aqui}} && \forall k \in \{0, \ldots, K\} \label{eq:initialwatervolumeinaquifer} \\
\dot{P}[k] &= \dot{P}^\ddagger[k] &&  \forall k \in \{0, \ldots, K\} \label{eq:precip}\\
\dot{V}_{SGR}[k] &= \dot{V}_{SGR}^\ddagger[k] && \forall k \in \{0, \ldots, K\} \label{eq:sgr}\\
\text{H}[k] - f_{H}(V_{\text{Mono}}[k]) &=0 && \forall k \in \{0, \ldots, K\} \label{eq:elevationfunction} \\
A_{\text{S}}[k] - f_{A_{\text{S}}}(V_{\text{Mono}}[k]) &= 0 && \forall k \in \{0, \ldots, K\} \label{eq:surfaceareafunction} \\
\lambda_{\text{Evap}}[k] - \lambda_{\text{FW}} \times \eta_{\bar{\rho}}[k] \times \eta_{\text{T}}[k] &= 0 &&  \forall k \in \{0, \ldots, K\}  \label{eq:evaporationrate} \\
\bar{\rho}[k] - \frac{(V_{\text{Mono}}[k] \times \rho_{\text{Water}} +\text{M}_{TDS})}{V_{\text{Mono}}{k} \times \rho_{\text{Water}}} &= 0  && \forall k \in \{0, \ldots, K\} \label{specificgravity} \\
\dot{V}_{P}[k] - \dot{P}[k] \times A_{\text{S}}[k] &=0 && \forall k \in \{0, \ldots, K\} \label{eq:inflowprecipvolume} \\
\dot{V}_{LA}[k] &=  \dot{V}^{\ddagger}_{LA} && \forall k \in \{0, \ldots, K\} \label{eq:inflowtoLA} \\
\dot{V}_{Perc}[k] - \lambda_{\text{Perc}}\times V_{\text{Mono}}[k] &= 0  &&  \forall k \in \{0, \ldots, K\}  \label{eq:percolationvolume} \\
\dot{V}_{\text{Evap}}[k] - \lambda_{\text{Evap}} \times A_{\text{S}}[k] &= 0 &&  \forall k \in \{0, \ldots, K\} \label{eq:evaporationasoutput}\\
\dot{V}_{FPDP}[k] - \dot{V}_{SGR}[k]- \dot{V}_{LA}[k] &= 0 &&  \forall k \in \{0, \ldots, K\}  \label{eq:FPDP_mathematics}\\
{T}[k] &= {T}^\ddagger[k] &&  \forall k \in \{0, \ldots, K\} \label{eq:temp} \\
\eta_{\text{T}}[k] - f_{T}({\text{T}}[k]) &=0 &&  \forall k \in \{0, \ldots, K\} \label{eq:effect of temperature}  \\
\eta_{\bar{\rho}}[k] - f_{\bar{\rho}}({\bar{\rho}}[k]) &= 0 &&  \forall k \in \{0, \ldots, K\}  \label{eq:effect of specific gravity} 
\end{align}

\noindent Each of these equations is described in turn:  
\begin{itemize}
\item Eqs. \ref{Eq:waterco} and \ref{Eq:watercontinuitylawforaquifer} show the water continuity law for the two stocks in the SD model.  They act as reservoirs receiving and discharging volumetric flow rates during a time interval $\Delta$T. In these equations, $\dot{V}_{\text{P}}$, $\dot{V}_{\text{FPDP}}$,   $\dot{V}_{\text{Dis}}$,  $\dot{V}_{\text{Evap}}$, $\dot{V}_{\text{Perc}}$ and $\dot{V}_{\text{GWWith}}$ are the volumetric flow rates associated with precipitation, flow past diversion point, groundwater discharge, evaporation, percolation, and groundwater withdrawal, respectively.
\item Eqs. \ref{eq:groundwaterdischarge} and \ref{eq:constantwithdrawl} represent the constant discharge of groundwater to the surface ($\dot{V}_{\text{Dis}}$) and the constant annual groundwater withdrawal ($\dot{V}_{\text{GWWith}}$) as exogenous variables, respectively.

\item Eqs. \ref{eq:initialwatervolumeinlake} and~\ref{eq:initialwatervolumeinaquifer} define the initial water volumes of the lake and the aquifer, respectively, at the start of the simulation period.
\item Eqs. \ref{eq:precip} and \ref{eq:sgr} show the time series of precipitation and Sierra Gauged Runoff exogenously defined in the SD model.
\item Eqs.~\ref{eq:elevationfunction} and~\ref{eq:surfaceareafunction} represent the elevation ($\text{H}[k]$) and surface area (${A}_{\text{S}}$) of the lake, respectively, as piece-wise constant functions of the lake volume.
\item Eq. \ref{eq:evaporationrate} calculates the evaporation coefficient ($\lambda_\text{Evap}$) which itself is a function of the Freshwater Evaporation Rate ($\lambda_{\text{FW}}$), the Specific Gravity Effect on Evaporation Rate ($\eta_{\bar{\rho}}$) and the Effect of Temperature on Evaporation ($\eta_{\text{T}}$).
\item Eq. \ref{specificgravity} calculates the specific gravity, where $\rho_{\text{Water}}$ is the density of water, and $M_{TDS}$ represents the total dissolved solids in the lake, assumed to be constant. 
\item Eq. \ref{eq:inflowprecipvolume} defines the calculation of volumetric inflow of precipitation over the lake surface area  as a function of ($\dot{P}$) and surface area ($A_{\text{S}}$).
\item Eq. \ref{eq:inflowtoLA} indicates the constant volumetric flow rate exported to LA ($\dot{V}_{LA}$) prior to entering the lake basin.
\item Eq. \ref{eq:percolationvolume} calculates the percolation volume for each time step, formalized as a portion of total water stored in the lake using the percolation coefficient ($\lambda_{\text{Perc}}$).
\item Analogous to the volumetric inflow from precipitation, Eq. \ref{eq:evaporationasoutput} delineates the computation of evaporation based on an evaporation coefficient ($\lambda_\text{Evap}$)(Eq.\ref{eq:evaporationrate}) which itself is a function of Freshwater Evaporation Rate ($\lambda_{\text{FW}}$), Specific Gravity Effect on Evaporation Rate ($\eta_{\bar{\rho}}$)(Eq. \ref{eq:effect of specific gravity}) and Effect of Temperature on Evaporation ($\eta_{\text{T}}$)(Eq. \ref{eq:effect of temperature}).
\item Eq. \ref{eq:FPDP_mathematics} calculates the volumetric flow rate of flow past diversion point.  In actuality, the volumetric inflow from Sierra Gauged Runoff is reduced by the constant volume exported to Los Angeles ($\dot{V}_{LA}$). 
\item Eq. \ref{eq:temp} defines time series of temperature modeled as an exogenous parameter.
\item Eqs. \ref{eq:effect of temperature} and~\ref{eq:effect of specific gravity} model the effects of temperature ($\eta_{\text{T}}$) and specific gravity ($\eta_{\bar{\rho}}$) on evaporation, respectively, using linear regression functions of temperature (${\text{T}}$) and specific gravity ($\bar{\rho}$).

\end{itemize}

\subsection{ Mono Lake in Model-Based Systems Engineering (MBSE)} \label{MonoSysML}
The SD model of the Mono Lake system is readily translated into an MBSE description stated in SysML.  Fig. \ref{fig:BDD} and \ref{fig:ACT} represent the form and the function of Mono Lake system as BDD and ACT Diagrams, respectively.  In Fig. \ref{fig:BDD}, the diagram frame denotes the system boundary and is labeled as a BDD of the `` Model" with ``Mono Lake system" as a model element and ``Mono Lake BDD" as the diagram name.   
\liinesbigfig{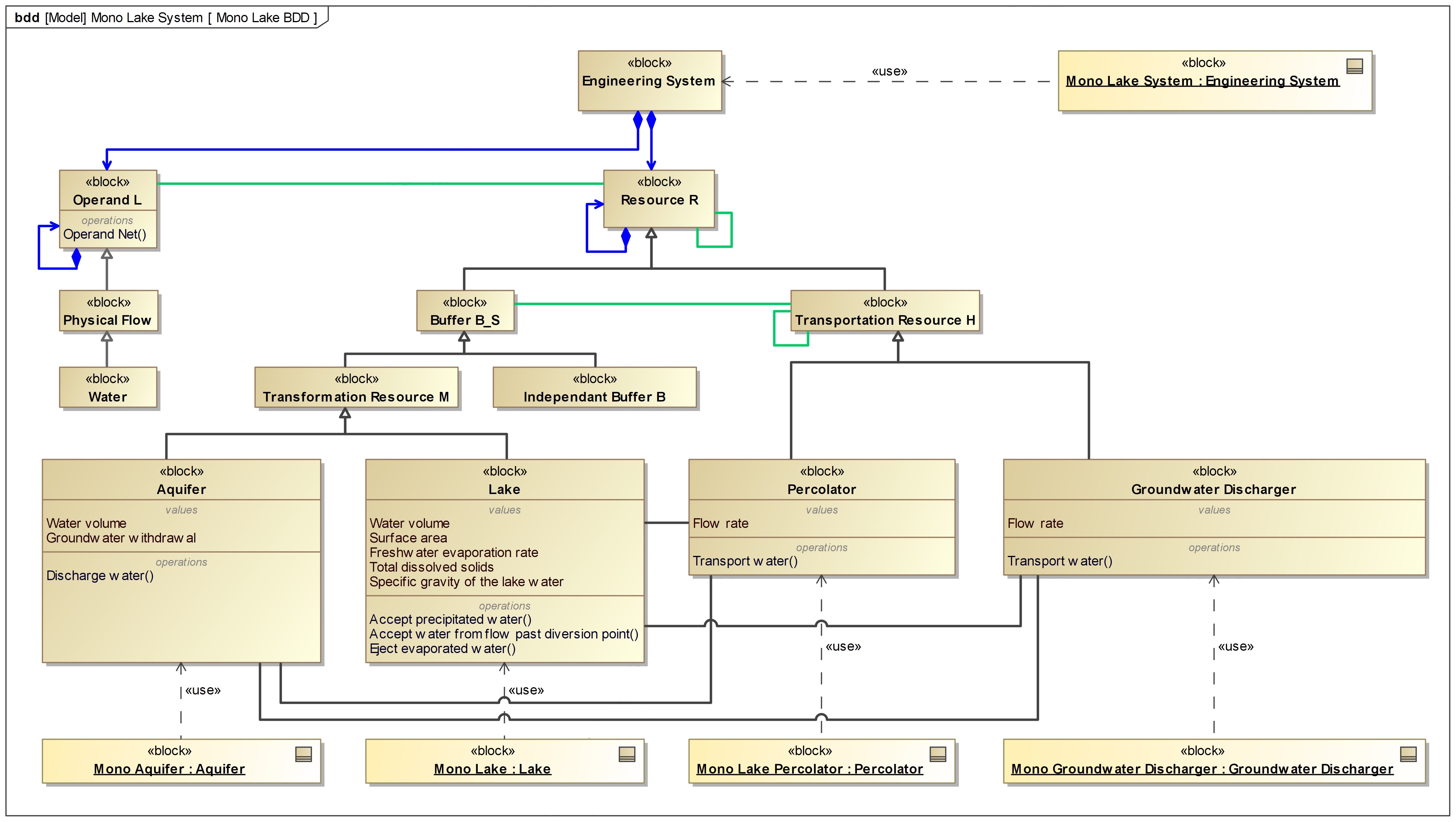}{Block Definition Diagram (BDD) of Mono Lake.}{fig:BDD}{7}
Fig. \ref{fig:BDD} uses the broad range of system thinking abstractions discussed in Sec. \ref{Sec:SDAssessment} to describe the Mono Lake system as an instance of an Engineering System in the HFGT meta-architecture.  Consequently, all of the parts of the Mono Lake system are also specializations of the parts of the HFGT meta-architecture.  Water is the only operand in the Mono Lake system.  Similarly, Mono Aquifer and Mono Lake blocks are instances of their associated classes which in turn are types of transformation resources $(M)$.  Additionally, the Mono Lake Percolator and the Mono Groundwater Discharger. More specifically, it displays as a formal specialization of the HFGT meta-architecture in Fig. \ref{fig:HFGT_BDD}.
 
As explained in Sec. \ref{Sec:MBSEHFGT}, the elements on the BDD are referred to as ``elements of definition" that are interrelated via either associations, generalizations or dependencies, which are three main types of relationships existing between blocks. As shown in Fig. \ref{fig:BDD}, ``engineering system" decomposes into operands $(L)$ and resources $(R)$, and resources are classified into buffer $(B_S)$ and transportation resource $(H)$. The operands are the objects of the verbs as explained in Sec. \ref{subsec:HFGT} (e.g., energy, water, oil). In the context of Mono Lake, Fig. \ref{fig:BDD} shows that the only operand is water as a type of physical flow that is being transported or transformed in the system \cite{thompson2024hetero}.  Among the primary blocks of the HFGT meta-architecture, buffers $(B_S)$ are classified into transformation resources $(M)$ and independent buffer $(B)$. Transformation resources $(M)$ perform transformation processes on operands (including bringing them across the system boundary). Mono Lake and Mono Aquifer are classified as transformation resources $(M)$, as they receive the operand (i.e., water) from outside the system boundary or discharge it outside the boundary. Transportation resources $(H)$ are another type of resource $(R)$ within the HFGT meta-architecture.  They are responsible for transporting operands from one buffer to another buffer. In the case of Mono Lake, the system has only two physical components:  the Mono Lake Percolator and the Mono Groundwater Discharger.  Both are classified as transportation resources $(H)$.

Fig. \ref{fig:BDD} also shows the value properties (i.e. attributes) and operations of each block component of the Mono Lake system .  The values associated with the inflows, outflows, the lake and the aquifer are noted as flow rate of water and water volume. Meanwhile, the operations on each block explain what each specific block does.  For example, Mono Lake accepts precipitated water from outside the system's boundary. In total, there are four blocks in the system which carry out six operations (or processes)(Fig. \ref{fig:BDD}). The values for Mono Lake as the main block in the BDD are all either the characteristics of the lake or the components of the system formalized using these characteristics. The absence of an arrowhead on either end indicates a ``reference association", allowing the blocks to interact or access each other for specific purposes through this connection.
In all, the HFGT  meta-architecture encompasses all the capabilities necessary to model the form of a complex system such as Mono Lake (Fig. \ref{fig:BDD}).

\liinesbigfig{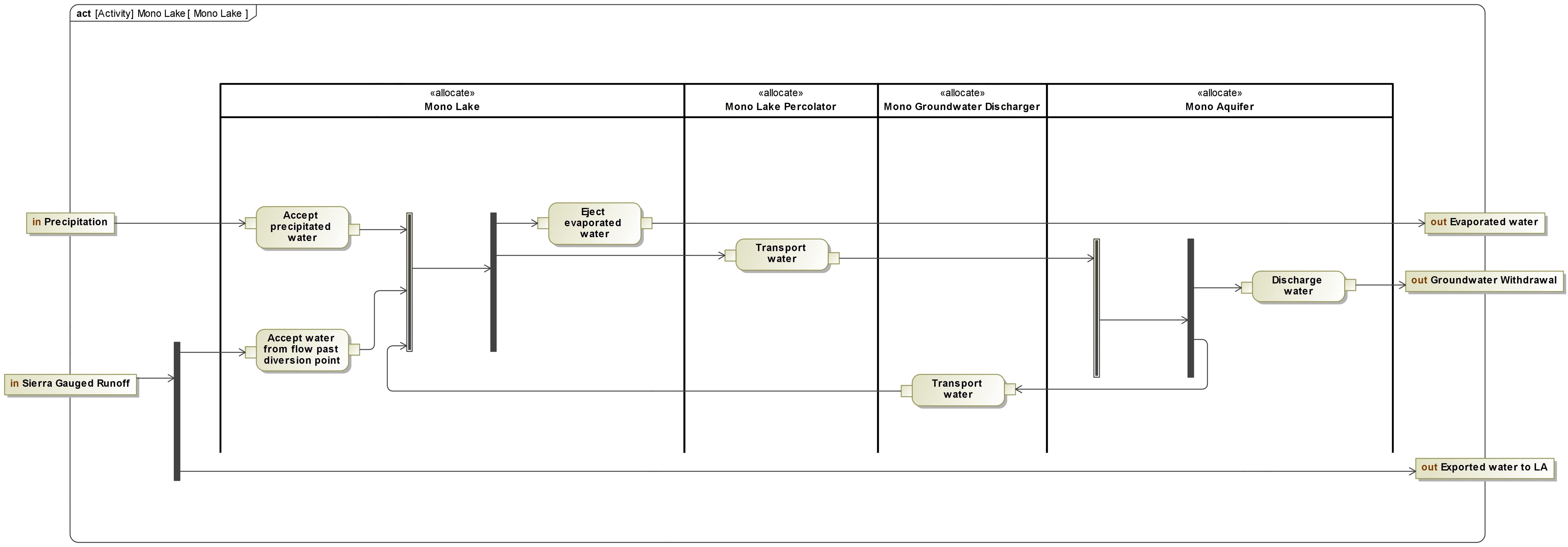}{ACT diagram of Mono lake showing the structural movement of objects throughout the entire system.}{fig:ACT}{7}

As explained earlier, compared to SD, ACT diagrams (e.g. Fig. \ref{fig:ACT}) offer greater readability and a more detailed behavior of the system by explicitly defining the boundary and components, distinguishing between the system's form and function, and representing its functionality using a broader range of abstractions. Specifically, when working with stakeholders, the complex structure of SD often prevents readers from fully grasping the nuances of the system, participating in the modeling processes, and trusting the accuracy and reliability of the model. As a result, the legitimacy, credibility and salience \cite{cash2020salience} of ACT would be considerably higher than those of SD in the eyes of stakeholders trying to learn or participate in the modeling process. As shown in Fig. \ref{fig:ACT}, variables such as ``Precipitation"  and ``Sierra Gauged Runoff" serve as exogenous input parameters within the system and are represented on the model's boundary as activity parameter nodes. Similarly, exogenous output variables, including ``Evaporated Water", ``Groundwater Withdrawal", and ``Exported Water to LA", denote outputs in the Mono Lake ACT. The placement of these exogenous variables indicates whether they are inputs or outputs.  By convention, all input parameters are positioned on the left boundary, while output parameters are located on the right boundary of the diagram.  

The ACT diagram has four swim lanes, Mono Lake, Mono Lake Percolator, Mono Groundwater Discharger and Mono Aquifer, that clarify the responsibilities of each of the resources in the system.  The Mono Lake swim lane, for instance, has three associated activities shown by rounded-corner boxes. In total, there are six activities that the four resources of the system are associated with, which is consistent with the six operations of the blocks in the BDD diagram (see Fig. \ref{fig:BDD} and  \ref{fig:ACT}). As an ontological rule, the activity should be written using human language as a ``verb + object" predicate. To guide readers when reading ACT diagrams, both the input and the output pins of every activity show the objects coming in and out of each activity.  This feature is essential when the system has many operands that it acts upon.  As a portion of water has been allocated to LA, the fork is located outside of the swimlanes, transporting water directly to LA.  In nearly all environmental systems, surface and subsurface reservoirs function as implicit storage units, storing objects such as water, sediments, and pollutants. Both the lake and aquifer exemplify this role in Mono Lake system, acting as implicit water storage reservoirs represented by horizontal joins sending water to the forks. The Mono Lake Percolator and Mono Groundwater Discharger swimlanes facilitate the bidirectional movement of water between the lake and the aquifer.

\subsection{Translating the SD model into  Hetero-functional Network Minimum Cost Flow (HFNMCF) Problem}
\label{HFNMCFmathematics}
This section translates the Mono Lake system equations presented in Sec. \ref{mathematicalequationsofmono} to the corresponding Hetero-functional Network Minimum Cost Flow (HFNMCF) problem.

\begin{align}
\text{minimize} \quad Z &= 0 \label{eq:objectivefunction}\\
-\begin{bNiceMatrix}
V_{Mono} \\ V_{Aqui}
\end{bNiceMatrix}[k+1] + 
\begin{bNiceMatrix}
V_{Mono} \\ V_{Aqui}
\end{bNiceMatrix}[k] +
\begin{bNiceMatrix}
1 & 1 & -1 & 0 & -1 & 1\\
0 & 0 &  0 & -1 & 1 & -1 
\end{bNiceMatrix}
\begin{bNiceMatrix}
\dot{V}_P \\ \dot{V}_{FPDP}\\ \dot{V}_{Evap}\\ \dot{V}_{GWWith}\\ \dot{V}_{Perc}\\ \dot{V}_{GDis}
\end{bNiceMatrix}[k] \Delta T
&= 0^{2\times 1} && \forall k \in \{1, \ldots, K\} \label{eq:statetransitionfunct}\\
\begin{bNiceMatrix}
0 & 0 & 0 & 1 & 0 & 0\\
0 & 0 & 0 & 0 & 0 & 1
\end{bNiceMatrix} 
\begin{bNiceMatrix}
\dot{V}_P \\ \dot{V}_{FPDP}\\ \dot{V}_{Evap}\\ \dot{V}_{GWWith}\\ \dot{V}_{Perc}\\ \dot{V}_{GDis}
\end{bNiceMatrix}[k] &= 
\begin{bNiceMatrix}
\dot{V}_{\text{GWWith}}^\ddagger\\ 
\dot{V}_{\text{GDis}}^\ddagger
\end{bNiceMatrix}[k] 
 && \forall k \in \{0, \ldots, K\} \label{eq:exogMono1}\\
\begin{bNiceMatrix}
1 & 0  \\
0 & 1  
\end{bNiceMatrix}
\begin{bNiceMatrix}
V_{Mono} \\ V_{Aqui}
\end{bNiceMatrix}[1] &= 
\begin{bNiceMatrix}
V_{Mono0}^\ddagger \\ V_{Aqui0}^\ddagger
\end{bNiceMatrix}  && \forall k \in \{1, \ldots, K\} \label{eq:initialConditions1}\\
\dot{P}[k] =& \dot{P}^\ddagger[k]  && \forall k \in \{0, \dots, K\} \label{eq:precipHFGTtimeseries}\\
\dot{V}_{SGR}[k] &= \dot{V}_{SGR}^\ddagger[k] && \forall k \in \{0, \dots, K\} \label{eq:SGR_HFGT}\\
H[k] - \left(0.0265 \times 
\begin{bNiceMatrix}
1 & 0
\end{bNiceMatrix}
\begin{bNiceMatrix}
V_{Mono} \\ V_{Aqui}
\end{bNiceMatrix}[k] + 6288.5\right) &= 0^{1\times 1} && \forall k \in \{0, \ldots, K\} \label{eq:elevationfunc1}\\
A_{\text{S}}[k] - \left(0.008 \times 
\begin{bNiceMatrix}
1 & 0
\end{bNiceMatrix}
\begin{bNiceMatrix}
V_{Mono} \\ V_{Aqui}
\end{bNiceMatrix}[k] + 15.44\right) &= 0^{1\times 1} && \forall k \in \{0, \ldots, K\} \label{eq:surfaceareafunction1}\\
\lambda_{Evap}[k] - \left(3.75 \times \eta_{\bar{\rho}}[k] \times \eta_{T}[k]\right) &= 0^{1\times 1} && \forall k \in \{0, \ldots, K\} \label{eq:Evaprate}\\
\bar{\rho}[k] - \frac{
\begin{bNiceMatrix}
1 & 0
\end{bNiceMatrix}
\begin{bNiceMatrix}
V_{Mono} \\ V_{Aqui}
\end{bNiceMatrix}[k] \times 1.36 + 250}{
\begin{bNiceMatrix}
1 & 0
\end{bNiceMatrix}
\begin{bNiceMatrix}
V_{Mono} \\ V_{Aqui}
\end{bNiceMatrix}[k] \times 1.36} &= 0^{1\times 1} && \forall k \in \{0, \ldots, K\} \label{specificgravityhf}\\
\begin{bNiceMatrix}
1 & 0 & 0 & 0 & 0 & 0
\end{bNiceMatrix}
\begin{bNiceMatrix}
\dot{V}_P \\ \dot{V}_{FPDP}\\ \dot{V}_{Evap}\\ \dot{V}_{GWWith}\\ \dot{V}_{Perc}\\ \dot{V}_{GDis}
\end{bNiceMatrix}[k] -  
\begin{bNiceMatrix}
\dot{P}[k]
\end{bNiceMatrix} \times A_{\text{S}}[k] &= 0^{1\times 1} && \forall k \in \{1, \ldots, K\} \label{eq:devicemodelofpercep1}\\
\begin{bNiceMatrix}
0 & 0 & 0 & 0 & 1 & 0
\end{bNiceMatrix}
\begin{bNiceMatrix}
\dot{V}_P \\ \dot{V}_{FPDP}\\ \dot{V}_{Evap}\\ \dot{V}_{GWWith}\\ \dot{V}_{Perc}\\ \dot{V}_{GDis}
\end{bNiceMatrix}[k] - 0.01 \times 
\begin{bNiceMatrix}
1 & 0
\end{bNiceMatrix}
\begin{bNiceMatrix}
V_{Mono} \\ V_{Aqui}
\end{bNiceMatrix}[k] &= 0^{1\times 1} && \forall k \in \{1, \ldots, K\} \label{eq:devicemodelofperc1}\\
\begin{bNiceMatrix}
0 & 0 & 1 & 0 & 0 & 0
\end{bNiceMatrix}
\begin{bNiceMatrix}
\dot{V}_P \\ \dot{V}_{FPDP}\\ \dot{V}_{Evap}\\ \dot{V}_{GWWith}\\ \dot{V}_{Perc}\\ \dot{V}_{GDis}
\end{bNiceMatrix}[k] - \lambda_{Evap}[k] \times A_{\text{S}}[k] &= 0^{1\times 1} && \forall k \in \{1, \ldots, K\} \label{eq:devicemodelofevap}\\
\begin{bNiceMatrix}
0 & 1 & 0 & 0 & 0 & 0
\end{bNiceMatrix}
\begin{bNiceMatrix}
\dot{V}_P \\ \dot{V}_{FPDP}\\ \dot{V}_{Evap}\\ \dot{V}_{GWWith}\\ \dot{V}_{Perc}\\ \dot{V}_{GDis}
\end{bNiceMatrix}[k] - \dot{V}_{SGR} - \dot{V}_{LA} &= 0^{1\times 1} && \forall k \in \{1, \ldots, K\} \label{eq:FPDP}\\
{T}[k] &= {T}^\ddagger[k] && \forall k \in \{0, \dots, K\} \label{eq:temperaturetimeseries}\\
\eta_{\text{T}}[k] - \left(0.06 \times \text{T}[k] - 0.08\right) &= 0^{1\times 1} && \forall k \in \{0, \ldots, K\} \label{eq:effect of temperature1}\\
\eta_{\bar{\rho}}[k] - \left(-0.9 \times \bar{\rho}[k] + 1.9\right) &= 0^{1\times 1} && \forall k \in \{0, \ldots, K\} \label{eq:effect of specific gravity1}\\
\dot{V}_{LA}[k] &= 16 && \forall k \in \{0, \ldots, K\} \label{eq:inflow_to_LA_HFNMCFP}
\end{align}

To elaborate:  
\begin{itemize}
\item Eq. \ref{Eq:ObjFunc} simplifies to Eq. \ref{eq:objectivefunction} because there are no defined objectives such as minimizing, maximizing or equilibrium on water flow, volume, or state of the operand (e.g., dissolved oxygen concentration) in the Mono lake system or aquifer. Consequently, a dummy objective function ($Z=0$) is defined. 
\item x[k]=$\left[ V_{\text{Mono}};\ V_{\text{Aqui}};\ \dot{V}_P;\ \dot{V}_{\text{FPDP}};\ \dot{V}_{\text{Dis}};\ \dot{V}_{\text{Evap}};\ \dot{V}_{\text{Perc}};\ \dot{V}_{\text{GWWith}} \right][k] \quad \forall k \in \{1, \ldots, K\}$ highlights the vector of primary decision variables of Mono Lake system.
\item Eq. \ref{Eq:ESN-STF1} simplifies to Eq. \ref{eq:statetransitionfunct}, which is a translation of Eq. \ref{Eq:waterco} and
\ref{Eq:watercontinuitylawforaquifer} from SD model. This equation captures the same underlying relationships in a more concise form. In this equation:
\begin{itemize}
\item $Q_B[k]$ is the volume of water in the buffers--the Mono lake and the Aquifer--$V_{Mono}[k]$ and $V_{Aqui}[k]$.
\item $MU[k]\Delta T = M^+U^+[k]\Delta T - M^-U^-[k]\Delta T$ given that $M$ is the set of arcs of the hetero-functional incidence matrices where:
\begin{align}
M^+ &= \begin{bmatrix} 1 & 1& 0 & 0 &0 & 1 \\ 0 & 0 & 0 & 0 & 1 & 0\end{bmatrix}\\
M^- &= \begin{bmatrix} 0 & 0& 1 & 0 &1 &0 \\ 0 & 0 & 0 & 1 & 0 & 1\end{bmatrix} \\
M = M^+ - M^- &= \begin{bmatrix} 1 & 1 & -1 &0 & -1 & 1\\ 0 & 0 &0 & -1 & 1& -1 \end{bmatrix}
\end{align}
While these hetero-functional incidence matrices can be calculated manually for small systems like Mono Lake, large complex systems-of-systems require automated means such as the openly available HFGT toolbox \cite{thompson2020hetero}.  
\end{itemize}
\item The duration of each capability (Defn. \ref{defn:capabilityCh7}) is instantaneous. Therefore, $k_{d\psi}=0\ \;  \forall k, \forall \psi$.  Therefore, Eq. \ref{Eq:DurationConstraint} collapses to $U^+[k] = U^-_{\psi}[k] = U[k] = [\dot{V}_P ; \dot{V}_{FPDP} ; \dot{V}_{Evap} ; \dot{V}_{GWWith} ; \dot{V}_{Perc} ; \dot{V}_{GDis}][k]$.

\item Eqs. \ref{Eq:ESN-STF2}, \ref{Eq:OperandNet-STF1},  \ref{Eq:OperandNet-STF2}, and \textcolor{magenta}{\ref{Eq:OperandNetDurationConstraint}} collapse to triviality given that the capacity of the transitions and the operand states are not defined in Mono Lake system. In other words, all transitions have a time duration of zero and the operand (i.e., water) does not change its state as it moves through the system. Consequently, these constraints are eliminated from the optimization problem entirely.
\item Eqs. \ref{Eq:SyncPlus} and \ref{Eq:SyncMinus} also collapse to triviality given that there are no synchronization constraints defined in the system that couple the input and output firing vectors of the engineering system net to the input and output firing vectors of the system operand nets, respectively. In simpler terms, because the operands don't change state, there is no need to syncronize the change of their state.  
\item Eq. \ref{CH6:eq:HFGTprog:comp:Bound} simplifies to Eq. \ref{eq:exogMono1}, which translates Eq. \ref{eq:groundwaterdischarge} and \ref{eq:constantwithdrawl} of the SD model, to represent the forcing outflow function from the aquifer. In fact, this equation specifies a boundary condition constraint that allows some of the engineering system net firing vector’s decision variables to be set to an exogenous constant. Note that the input and output firing vectors have been consolidated into a single firing vector because the transition durations are instantaneous.  
\item Eq. \ref{Eq:OperandRequirements} is also not required because there is no operand net to track the state of the water in the system.  
\item Eq. \ref{CH6:eq:HFGTprog:comp:Init} simplifies to Eq. \ref{eq:initialConditions1}, which is the translation of Eq. \ref{eq:initialwatervolumeinlake} and \ref{eq:initialwatervolumeinaquifer} from the SD model, to define the initial volume and mass in Mono lake and the aquifer.
\item Eq. \ref{CH6:eq:HFGTprog:comp:Fini} is a final condition constraint and is eliminated because the simulation reflects an initial value problem/simulation. 
\item Eq. \ref{ch6:eq:QPcanonicalform:3} are eliminated by relaxation.  In this system, no lower or upper bounds are placed on the primary decision variables. Consequently, the inequality constraints on primary decision variables in Eq. \ref{ch6:eq:QPcanonicalform:3} are set to positive and negative infinity. 
 $\underbar{E}_{CP} \rightarrow -\infty$, $\overline{E}_{CP} \rightarrow \infty$. 
\item  Eq. \ref{Eq:DeviceModels}, describing domain-specific device models, is transformed to Eqs. \ref{eq:precipHFGTtimeseries}, \ref{eq:SGR_HFGT}, \ref{eq:elevationfunc1}, \ref{eq:surfaceareafunction1}, \ref{eq:Evaprate} and \ref{specificgravityhf}.  These correspond to Eqs. \ref{eq:precip}, \ref{eq:sgr}, \ref{eq:elevationfunction}, \ref{eq:surfaceareafunction}, \ref{eq:evaporationrate}, and \ref{specificgravity} respectively in the SD model.  Additionally, Eqs. \ref{eq:devicemodelofpercep1}-\ref{eq:FPDP} correspond to the constitutive laws of the Mono Lake system from the SD model in  Eqs. \ref{eq:inflowprecipvolume}, \ref{eq:percolationvolume}, \ref{eq:evaporationasoutput}, and \ref{eq:FPDP_mathematics} respectively.  
\item Eq. \ref{Eq:DeviceModels2} is transformed into Eqs. \ref{eq:temperaturetimeseries}, \ref{eq:effect of temperature1}, \ref{eq:effect of specific gravity1} and \ref{eq:inflow_to_LA_HFNMCFP} demonstrating the device models corresponding to the auxiliary variables shown as Eqs. \ref{eq:temp}, \ref{eq:effect of temperature},\ref{eq:effect of specific gravity} and \ref{eq:inflowtoLA} respectively in the SD model.
\end{itemize}

\section{ Results and Discussion}
\label{Sec:DiscussionandResults}

The Mono Lake system example demonstrates that HFGT is not only capable of quantitatively replicating the SD model, but also offers a valuable tool for modeling  systems with considerably greater complexity. This section describes the SD and HFGT simulations in terms of their exogeneous parameters, compares their simulation results, and discusses the potential benefits that HFGT may provide for the Anthropoecene systems-of-systems; especially when they exhibit greater size and complexity.  

The SD and HFGT models were simulated with the exogeneous parameters depicted in Fig. \ref{MonoOnlySD}.  Both models were calibrated and simulated on an annual time step to forecast the hydrological dynamics of the Mono Lake and the Mono Aquifer over the period 1990–2090. Fig. \ref{MonoOnlySD} presents the temporal evolution of key climatic and hydrological drivers of Mono Lake system, including precipitation, groundwater withdrawal, temperature, and Sierra Gauged Runoff. Except for the constant annual groundwater extraction rate of 6,800 KAF from the Mono Aquifer, future trajectories for precipitation, temperature, and Sierra runoff, are generated using a Markov Chain approach based upon historical data spanning 1960 to 1990.

\liinesbigfig{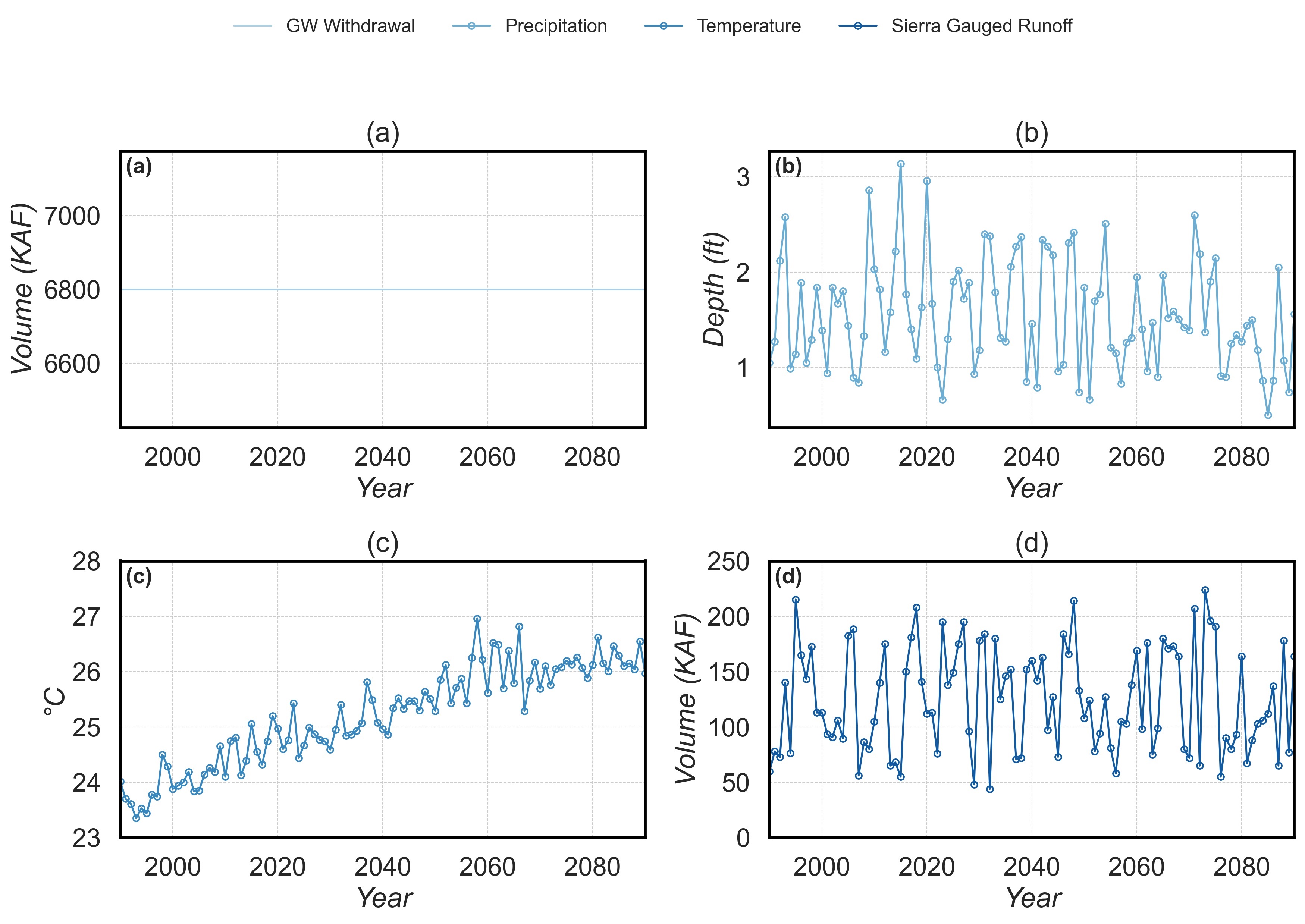}{Exogeneous parameters used in SD and HFGT simulations for 1990-2090: (a) groundwater withdrawal from Mono Aquifer, (b) precipitation, (c) temperature and (d) Sierra Gauged Runoff.}{MonoOnlySD}{7}

Figs. \ref{HFGT_SD} and \ref{SD:HFGT 2} show that when the Mono Lake system is simulated via the SD and HFGT models, they yield nearly identical simulation results, indicating a high degree of consistency between the two modeling approaches. In Fig. \ref{HFGT_SD}, the graph shows that both the SD and HFGT simulate a decline of approximately 728 KAF in the Mono Lake water volume by the end of 2090 which is mainly due to evaporation, reduced levels of precipitation and exporting water to LA.  The Mono Aquifer is also simulated to experience a decrease in water volume of about 680 KAF by both modeling approaches, primarily due to reduced percolation and constant extraction from wells to meet demands. The trajectories produced by both approaches are nearly indistinguishable, as shown in Figs. \ref{HFGT_SD} and \ref{SD:HFGT 2}. The normalized Root Mean Square Error (nRMSE) between the two approaches for both the Mono Lake and the Mono Aquifer is approximately 0.15\% and 0.000\%, respectively.  The negligible differences in results are likely attributed to numerical implementation such as the numerical solver, time-step discretization, and error propagation mechanisms.  

\liinesbigfig{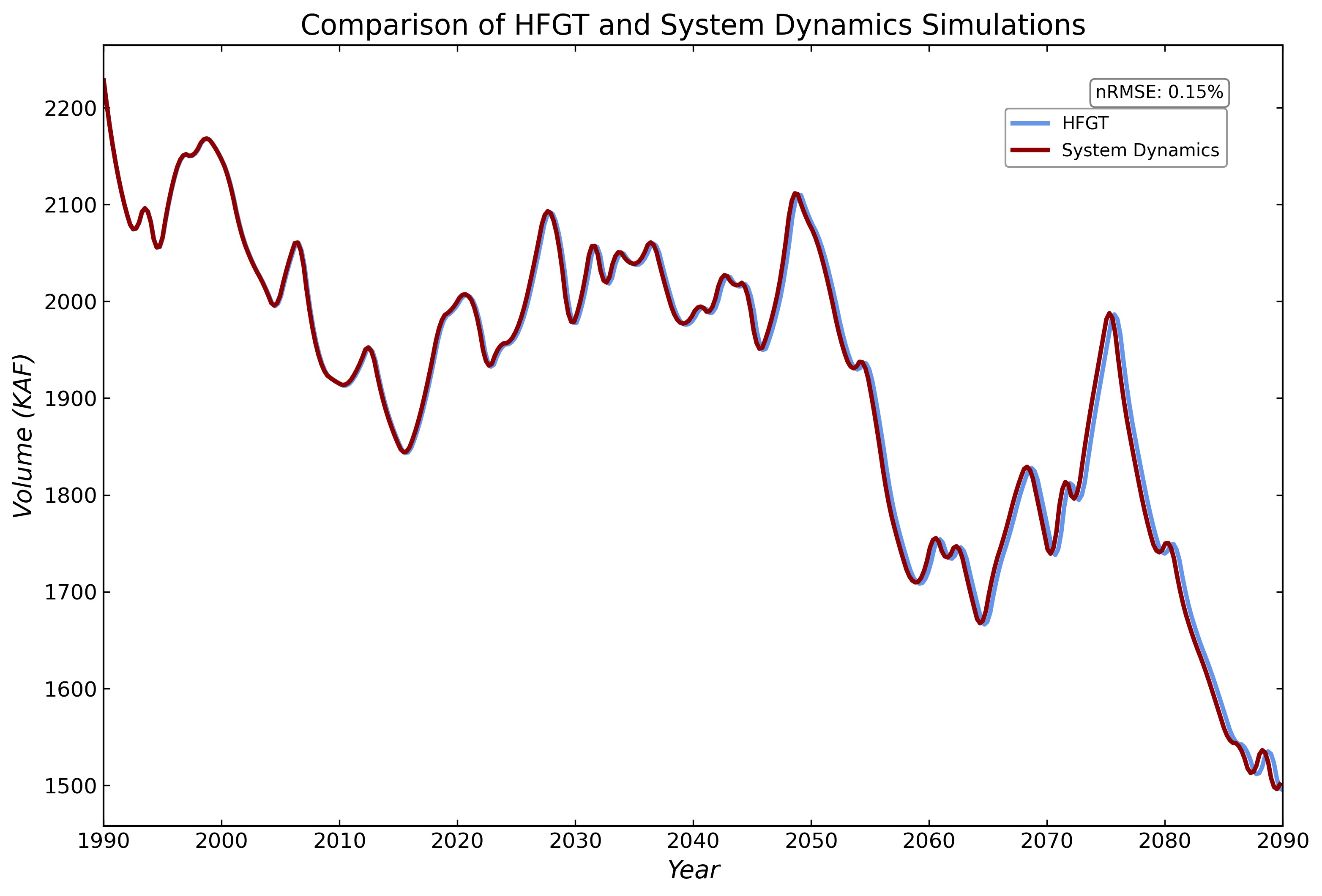}{Simulated Mono Lake Volume by SD and HFGT.}{HFGT_SD}{4.2}

\liinesbigfig{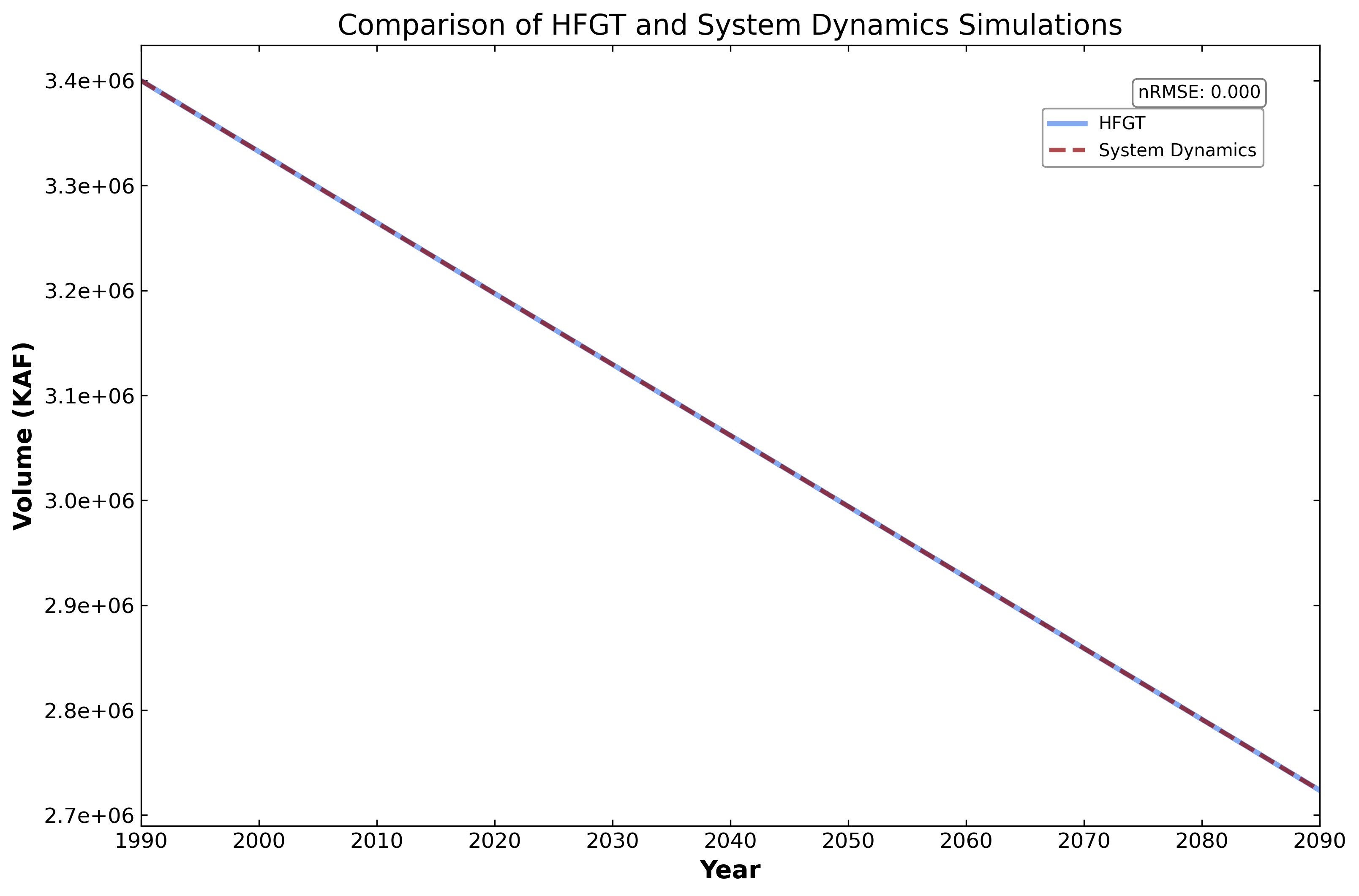}{Simulated Mono Aquifer Volume by SD and HFGT.}{SD:HFGT 2}{4.2}

Although the SD and HFGT models yield nearly identical simulation results, the HFGT model presents several benefits for Anthropocene systems of systems.
SD employs a relatively limited set of modeling primitives to separately show what the system is made up of and how it functions.These limitations become apparent when modeling multi-domain systems that demand detailed representation of system entities, variable types, interdependencies, and structural dynamics. It also struggles to manipulate heterogeneous components, structures, processes, interactions, and scales. Handling large datasets associated with system elements and processing it for multiple interconnected systems is another obstacle that SD cannot overcome due to the intrinsic limitations of the modeling software. Additionally, the SD software lacks the capacity to handle the vast data demands imposed by diverse mathematical formulations including differential-algebraic equations, discrete-event processes, stochastic differential equations, optimization-based constraints, and hybrid systems combining continuous and discrete dynamics. MBSE stated in SysML, on the other hand, offers a nuanced representation of systems boundary, form, and function, capturing the fundamental or inherent nature of heterogeneous entities and their interdependencies. The significance of explicit representation of Anthropocene system-of-systems becomes more evident in a participatory modeling setting with multiple stakeholders\cite{hood2021chesapeake,hemmerling2020elevating} to facilitate participatory modeling and include diverse knowledge sources.

Furthermore, using such a shared common language like SysML serves to reconcile diverse knowledge sources and facilitate the representation and readability of models of coupled systems-of-systems\cite{podesta2013interdisciplinary,videira2009scoping}.  The application of HFGT enables the quantification of models represented in SysML. The heterogeneity of these models encompasses 
algebraic, differential, and differential algebraic equations; be they linear or nonlinear.  SysML also supports deterministic and stochastic models with continuous and discrete variables.  The majority of these modeling options remain unsupported by SD software. Together, the many features of SysML and HFGT make for an effective  computational approach that is well suited to the diverse characteristics of Anthropocene system-of-systems.  Ultimately, the choice between SD and MBSE-HFGT depends fundamentally on the modeling objective.  While SD is well-suited for representing and simulating fairly simple systems with limited structural complexity, MBSE-HFGT offers a more scalable and expressive framework capable of capturing and simulating the intricacies and interdependencies found in highly complex Anthropocene systems.  

\section{ Conclusion and future work}\label{Sec:Conclusion}
\subsection{Conclusion}

MBSE, SysML and HFGT are proposed as an alternative modeling and analysis framework to SD. Although both SD and the proposed MBSE-SysML-HFGT framework provide robust modeling approaches, the main difference lies in their ability to represent complex system structures.  MBSE and the associated SysML diagrams (e.g., BDD and ACT diagrams) offer greater readability and representation of detailed structures and behaviors.  They explicitly define the system's boundary and components, distinguish between system form and function, and represent the system functionality using a broader range of systems thinking abstractions.  In contrast, SD uses a fairly limited number of abstractions, representing system form and function in a single graphical model without an explicit depiction of the system boundary. While SysML provides a common modeling language for large complex systems of systems, HFGT provides an analytical means of translating graphical SysML models into mathematical and computational models. In addition, HFGT is conceptually rooted in the universal structure of human language providing the basis for a discipline-agnostic ontology that facilitates cross-disciplinary applications. The two modeling approaches are demonstrated on the Mono Lake system, a hydrological system that has been previously studied with system dynamics. The demonstration confirms that the MBSE-SysML-HFGT framework can reproduce the modeling and simulation results developed through an SD approach, showing that no analytical insight is lost compared to SD. Indeed, the MBSE-SysML-HFGT framework provides greater analytical insight for the same illustrative system.

\subsection{Future work}
To integratively address the interdependent societal challenges of the Anthropocene, a system-of-systems (SoS) convergence paradigm that includes an SoS computational framework, an SoS decision-support system, and an SoS educational pedagogy is needed \cite{little2023earth}. Unfortunately, because of the large number of systems involved, it is essentially impossible to integrate models of all the systems directly due to their diverse, discipline-specific ontologies. The MBSE-SysML-HFGT framework can be used to overcome these limitations, first translating “real-world” Anthropocene systems into SysML and then using HFGT to algorithmically traverse the gap from the graphical SysML model to the associated mathematical model, and ultimately to the computational model, with an HFGT toolbox available to perform the calculations.

The SoS convergence paradigm must also include the central role of human decision-making in convergent Anthropocene systems, with a comprehensive SoS decision-support system that provides a structured approach for translating SoS computational results into real-world insights for diverse stakeholders. Finally, there is an urgent need for an SoS educational pedagogy to train a new generation of Anthropocene system integrators (e.g., students, academics, practitioners, and stakeholders) with knowledge of the convergence paradigm. They need to conceptualize Anthropocene societal problems holistically, think coherently in terms of common ontological constructs, produce and simulate SoS computational models, facilitate decision-support conversations filled with coherent, real-world insights, and use these insights to once again conceptualize societal problems.

The SoS convergence paradigm is currently being implemented and validated for three interdependent societal challenges (eutrophication, agricultural impacts, and economic growth) in the Chesapeake Bay Watershed. The Chesapeake Bay Program (CBP) operates a sophisticated watershed management system balancing competing priorities across multiple federal, state and private institutions. Building on decades of CBP research developing models of land-use, watershed and estuary systems \cite{hood2021chesapeake}, a model of an economic system as well as a governance system are being added. An agile approach is being adopted to coherently integrate the land use, watershed and economic systems in a first phase, followed by estuary and governance systems in a second phase. SysML, the common graphical and spoken modeling language, provides the foundation for the entire SoS convergence paradigm, including the computational framework, decision-support system, and educational pedagogy, ensuring convergence across the multiple disciplinary ontologies of the five integrated systems.

\bibliographystyle{IEEEtran}
\bibliography{LIINESLibrary,LIINESPublications,references}

\end{document}